\documentclass[aps,pre,superscriptaddress,showpacs,preprintnumbers,preprint]{revtex4-1}
\usepackage{epsfig}
\usepackage{graphicx}
\usepackage{subfigure}
\usepackage{float}
\usepackage{multirow}
\usepackage{amssymb}
\usepackage{amsmath}
\usepackage{picinpar}

\begin{document}
\newcommand{\figwidth}{0.65\textwidth}
\newcommand{\figwidthgnu}{0.55\textwidth}
\newcommand{\subfigwidth}{0.45\textwidth}
\newcommand{\subfigwidthgnu}{0.33\textwidth}
\newcommand{\figvspace}{\vspace{0.6cm}}
\newcommand{\fighspace}{\hspace{0.5cm}}

\newcommand{\feru}{F\lowercase{e}$_{1-\lowercase{x}}$R\lowercase{u}$_{\lowercase{x}}$}
\newcommand{\FeAl}{F\lowercase{e}$_{1-\lowercase{x}}$A\lowercase{l}$_{\lowercase{x}}$}
\newcommand{\FeMn}{F\lowercase{e}$_{1-\lowercase{x}}$M\lowercase{n}$_{\lowercase{x}}$}
\newcommand{\FeMnAl}{F\lowercase{e}$_{1-\lowercase{y}-\lowercase{x}}$M\lowercase{n}$_{\lowercase{y}}$A\lowercase{l}$_{\lowercase{x}}$}
\newcommand{\FeZnF}{F\lowercase{e}$_{1-\lowercase{x}}$Z\lowercase{n}$_{\lowercase{x}}$F$_2$}
\newcommand{\Fe}{F\lowercase{e}}
\newcommand{\Ru}{R\lowercase{u}}
\newcommand{\Co}{C\lowercase{o}}
\newcommand{\Ni}{N\lowercase{i}}
\newcommand{\T}{\bar{T}}

\newcommand{\smin}{\mbox{\scriptsize{min}}}
\newcommand{\smax}{\mbox{\scriptsize{max}}}
\newcommand{\stot}{\mbox{\scriptsize{tot}}}
\newcommand{\ms[2]}{\mbox{\scriptsize{#2}}} 
\newcommand{\im}{i}
\newcommand{\jm}{j}
\newcommand{\sumnn}{\sum_{\langle\im,\jm\rangle}}
\newcommand{\sumsig}{\sum_{\{\sigma\}}}

\title{Magnetic properties and critical behavior of disordered
F\lowercase{e}$_{1-\lowercase{x}}$R\lowercase{u}$_{\lowercase{x}}$
alloys: a Monte Carlo approach}

\author{I. J. L. Diaz}\email{ianlopezdiaz@gmail.com}
\affiliation{Universidade Federal da Fronteira Sul,
Campus Cerro Largo, 97900-000, Cerro Largo, RS, Brazil}
\author{N. S. Branco}\email{nsbranco@fisica.ufsc.br}
\affiliation{Departamento de F\'{i}sica,
Universidade Federal de Santa Catarina,
88040-900, Florian\'{o}polis, SC, Brazil}

\date{\today}

\begin{abstract}
We study the critical behavior of a quenched random-exchange Ising model with
competing interactions on a bcc lattice. This model was introduced in the study of the magnetic behavior of
F\lowercase{e}$_{1-\lowercase{x}}$R\lowercase{u}$_{\lowercase{x}}$
alloys for ruthenium concentrations $x=0\%$, $x=4\%$, $x=6\%$, and $x=8\%$.
Our study is carried out within a Monte Carlo approach, with the aid of
a re-weighting multiple histogram technique.
By means of a finite-size scaling analysis of several thermodynamic quantities,
taking into account up to the leading irrelevant scaling field term,
we find estimates of the critical exponents $\alpha$, $\beta$, $\gamma$, and $\nu$,
and of the critical temperatures of the model.
Our results for $x=0\%$ are in excellent agreement with those for the
three-dimensional pure Ising model in the literature.
We also show that our critical exponent estimates for the disordered cases
are consistent with those reported for the transition line between paramagnetic
and ferromagnetic phases of both randomly dilute and $\pm J$ Ising models.
We compare the behavior of the magnetization as a function of temperature with
that obtained by Paduani and Branco (2008), qualitatively confirming the mean-field result.
However, the comparison of the critical temperatures obtained in this work with
experimental measurements suggest that the model (initially obtained in a
mean-field approach) needs to be modified.
\end{abstract}


\maketitle

\section{Introduction}\label{introduction}
We study the magnetic properties and critical behavior of a quenched random-exchange Ising model,
through extensive Monte Carlo simulations.
Our starting point was the model proposed in Ref. \onlinecite{nilton:feru},
consisting of iron and ruthenium atoms randomly distributed in a body-centered cubic (bcc) lattice,
with probabilities $1-x$ and $x$, respectively.
In this model, atoms are treated as Ising spins,
\Fe-\Fe{} bonds are ferromagnetic (FM), with exchange integral $J$, while \Fe-\Ru{} and \Ru-\Ru{}
interactions are antiferromagnetic (AF) with exchange integrals $-\lambda J$ and $-\xi J$, respectively,
where $\lambda$ and $\xi$ depend on the ruthenium concentration, $x$, as follows: $\lambda\equiv\xi=\xi_0-\xi_1x$.
The parameters $\xi_0$ and $\xi_1$ were determined from the experimental values for the critical temperatures of the system
for some concentrations of ruthenium, reported in Ref. \onlinecite{artigo:paduani}, by fitting these data
to the mean-field solution of the model.

The main goal of this work is to establish the universality class of the spin model introduced in Ref. \onlinecite{nilton:feru},
through extensive numerical calculations of some of its thermodynamic quantities.
We used the metropolis algorithm to generate the data  and employed
re-weighting techniques and finite-size scaling (FSS) analysis.
We also compare the behavior of thermodynamic quantities
obtained by Monte Carlo simulations with experimental and mean-field results. 
We compare the values of $T_c(x)$ with the experimental measurements \cite{artigo:paduani} 
and theoretical estimates \cite{nilton:feru} to determine if the model for the \feru{} alloys
remains adequate in a non-mean-field approach. In this later work,
the parameters of the proposed Hamiltonian were obtained through a fit of the theoretical critical temperatures,
obtained in a mean-field-like approximation, with the measured values reported in Ref. \onlinecite{artigo:paduani}. The mean-field
result is obtained using the Bogoliubov inequality, in a similar fashion as the one employed in the study of order-disorder
transitions through a cluster variational method \cite{kikuchi1980}.

An important aspect of this work is the FSS analysis to determine the critical exponents. 
The study of this model is mainly an investigation of the effects of quenched disorder
on the critical behavior of a system of Ising spins, like so many theoretical and experimental
examples presented throughout the last three decades or more. 
An interesting fact of these disordered systems is the theoretical prediction 
that the introduction of disorder must change the universality class of the system if $\alpha>0$, 
where $\alpha$ is the specific heat critical exponent for the pure system. 
This result is known as the Harris criterion \cite{livro:cardy}. 
The $\alpha>0$ condition is satisfied for the case of the pure Ising model in three dimensions,
so we can expect three-dimensional disordered Ising models to belong to another universality class,
when compared to their uniform counterparts.
However, the Harris criterion says nothing about the new universality class.
Renormalization-group arguments infer that the universality class of the disordered three-dimensional Ising model
does not depend on the concentration $x$.
Moreover, experimental results show that, for small concentrations $x$ of AF impurities
(as in \FeMn{}, \FeZnF{} and \feru{} alloys) or nonmagnetic impurities (as in \FeAl{} alloys),
these systems have a continuous transition between a ferromagnetic and a paramagnetic phase
at a temperature $T_c(x)<T_c(x=0\%)$, with a critical behavior clearly different from the case without disorder ($x=0\%$),
but that seems to be independent of the concentration $x$, within the experimental precision \cite{belanger2000}.
By small concentrations we mean $x<x_c$ for nonmagnetic impurities, where $x_c$ is the critical concentration
above which there is only a single paramagnetic phase \cite{hasenbusch2007uc}.
In the case of AF impurities, we mean $x<x_g$, where $x_g$ is the concentration for which the transition line between
paramagnetic and ferromagnetic phases meets the transition line between paramagnetic and spin-glass phases,
where frustration is relevant \cite{livro:barkema}.

It is curious that, contrary to experimental results,
Monte Carlo simulations presented a wide range of different values for the critical exponents
of these disordered systems, all seeming to indicate some sort of dependence between universality class and impurity concentration.
Only recently (late 1990's onward), with the growth of processor capacity and available resources for simulations,
is theoretical research heading toward a solution to this apparent inconsistency,
and studies have shown that critical exponents are independent of impurity concentration
along the transition line between paramagnetic and ferromagnetic phases in disordered Ising
models \cite{ballesteros1998, calabrese2003tdr, hasenbusch2007uc, hasenbusch2007+-J}. 
These works have shown the importance of a careful analysis of the scaling behavior of thermodynamic functions, 
taking into account finite-size correction terms due to irrelevant fields in the Hamiltonian,
for a correct assessment of the critical exponents of these systems.

At first glance, one could suspect that our model does not belong to any of the previously discussed
universality classes. However, we could think of it as a member of
a more general family of models with bonds $+J$ and $-rJ$ randomly distributed with
probabilities $1-p$ and $p$ respectively.
We obtain the Ising model with randomly diluted bonds (RBIM) as $r=0$, the Ising model with $\pm J$ random
bonds for $r=1$ and our model for some combinations of $r$ and $p$.
Note also that the \feru{} system we study in this work has a crossover to the
Ising Model with randomly diluted sites (RSIM) when $x=10\%$.
Therefore, we make the hypothesis that our model belongs to the RSIM-RBIM universality class, as is the case also for
the paramagnetic-ferromagnetic transition line of the $\pm J$ Ising model \cite{hasenbusch2007uc, hasenbusch2007+-J}.
In this sense, our \feru{} system would be somewhere in between the $r=0$ and $r=1$ limits.
To assess this possibility, we have performed extensive Monte Carlo simulations, with the important
inclusion of correction-to-scaling terms, as discussed in the previous paragraph, in order to
estimate the critical exponents and critical temperatures of the model introduced in Ref. \onlinecite{nilton:feru}.
This model is explained in Sec. \ref{model}, the simulation and data analysis methods are discussed in Secs.
\ref{monte_carlo} and \ref{data_analysis} respectively and our results are presented and discussed in Sec. \ref{results}.

\section{Model}\label{model}
Ref. \onlinecite{artigo:paduani} shows that \feru{} alloys are found in the bcc structure for $x<30\%$ whereas, for $x\approx 30\%$,
the system undergoes a crystallographic transition to an hcp structure.
While in the bcc structure, the lattice parameter increases steadily with the Ruthenium concentration $x$ and
the system has a ferromagnetic-paramagnetic phase transition at $T_c(x)$, as seen in Tab. \ref{tab:Tc(x)}.
In Ref. \onlinecite{nilton:feru}, the authors present a detailed discussion
of experimental results and first-principle electronic-structure calculations that provide evidence that
ferromagnetic \Fe-\Fe{} bonds and antiferromagnetic \Fe-\Ru{} and \Ru-\Ru{} bonds should effectively model the \feru{} system.

%%%%%%%%%%%%%%%%%%%%%
%%% TAB
\begin{table}[h]\caption{Experimental values of critical temperatures for \feru{} alloys, taken from Ref. \onlinecite{artigo:paduani}.}
\begin{tabular}{cc}
	\hline\hline
	$x$ & $T_c ~(K)$\\
	\hline
	$0\%$  & 1043\\
	$2\%$  & 968(2)\\
	$4\%$  & 928(2)\\
	$6\%$  & 908(2)\\
	$10\%$ & 838(2)\\
	\hline\hline
\end{tabular}
\label{tab:Tc(x)}
\end{table}
%%%%%%%%%%%%%%%%%%%%%

The model proposed by Paduani and Branco (2008)\nocite{nilton:feru} consists of
\Fe{} and \Ru{} atoms randomly distributed on a bcc lattice with probabilities $1-x$ and $x$ respectively.
Each atom has a magnetic degree of freedom that is assumed to behave as an Ising-like spin,
so the system is described by a spin-1/2 Ising Hamiltonian
\begin{equation}\label{Ising:Hamiltonian}
	\mathcal{H}=
	-\sum_{\left\langle \im,\jm\right\rangle}
	J_{\im\jm}\sigma_\im\sigma_\jm,
\end{equation}
where the sum goes over all nearest-neighbor pairs, $\sigma_\im=\pm 1$ for all sites $\im$, and the exchange integral
$J_{\im\jm}$ takes the values $J$ for \Fe-\Fe{} bonds, $-\lambda J$ for \Fe-\Ru{} bonds, and $-\xi J$ for \Ru-\Ru{} bonds.

A mean-field solution \cite{nilton:feru} of Hamiltonian (\ref{Ising:Hamiltonian}),
using the Bogoliubov inequality \cite{livro:callen, livro:julia}, yields
\begin{equation}\label{eq:fit}
	\left\{\frac{(1-x)^2}{1+\exp(-2J/k_BT_c)}
	+\frac{2x(1-x)}{1+\exp(2\lambda J/k_BT_c)}
	+\frac{x^2}{1+\exp(2\xi J/k_BT_c)}\right\}=\frac{4}{7}.
\end{equation}
In Ref. \onlinecite{nilton:feru}, the authors assume that $\lambda=\xi\equiv \xi_0-\xi_1 x$
and present a least-square fit of Eq. (\ref{eq:fit})
to the experimental values of $T_c$ in Tab. \ref{tab:Tc(x)}.
The values reported for the parameters are $\xi_0=0.54(2)$ and $\xi_1=5.4(4)$,
so that the final form of the Hamiltonian reads:
\begin{equation}\label{Ising:feru:hamiltoniano}
	\mathcal{H}=
	-J\sum_{\left\langle \im,\jm\right\rangle}
	\zeta_{\im\jm}\sigma_\im\sigma_\jm
\end{equation}
where
\begin{equation}
\zeta_{\im\jm}=
\left\lbrace
	\begin{array}{lll}
		\zeta_{\ms{FeFe}} & = & 1 \\
		\zeta_{\ms{FeRu}} & = & -(0.54-5.4x) \\
		\zeta_{\ms{RuRu}} & = & -(0.54-5.4x),
	\end{array}
\right.
\end{equation}
and the spin variables $\sigma_\im$ take the values $\pm 1$
and the sum goes over all nearest-neighbor pairs.

\section{Monte Carlo Simulations}\label{monte_carlo}
We studied the three-dimensional Ising system described by the Hamiltonian
(\ref{Ising:feru:hamiltoniano}) within a Monte Carlo (MC) approach.
We have employed the metropolis algorithm \cite{artigo:metropolis, livro:barkema}
to simulate bcc lattices with $2L^3$ sites, periodic boundary
conditions, and several system sizes $L$, ranging from 5 to 50.
Each site on the lattice is randomly chosen to be an \Fe{} or \Ru{} atom with
probabilities $1-x$ and $x$, respectively.
All random numbers were generated using a Tausworthe (shift-register)
generator \cite{artigo:shift-register} with ``magic numbers''
$p=1279$ and $q=1063$ \cite{artigo:landau,shan-ho:private}.

A Monte Carlo step per spin (MCS) corresponds to $N=2L^3$ attempts to flip a single spin,
chosen at random, or a full sweep on the lattice, when
we perform sequential updates.
As the sequential algorithm proved to be far more efficient in generating
independent states than the more traditional random choice, it was
the one used for most of our simulations.
The comparison between these two procedures is presented in Subsec. \ref{ran_seq}.

Our largest simulations ran up to $4.5\times 10^7$ MCSs and we made sure to generate
at least $n=8000$ uncorrelated states, with $n$ given by
\begin{equation}
	n=\frac{n_{\ms{MCS}}-t_{\ms{eq}}}{2\tau},
\end{equation}
where $n_{\ms{MCS}}$ is the number of MCSs, $t_{\ms{eq}}$ is the equilibration time and $\tau$
is the largest correlation time estimated by the integral method \cite{livro:barkema}.
As it is the case for the metropolis dynamics, the largest correlation time
is, in general, obtained from the magnetization autocorrelation function.
Therefore, in our case, $\tau$ refers to the magnetization correlation time.

For practical reasons concerning data storage, we have only saved simulation data
every 10 MCS. However, we have rescaled our dynamic calculations and present all
correlation times in units of one MCS.
This rescaling introduces a systematic error on the integral correlation
time when typical values of $\tau$ are close to 10 MCS or smaller, as shown in Sec. \ref{ran_seq}.
This error is due to the numerical integration using the trapezoid rule,
as the integral of an exponential decay approaches the exact value from above
with numerical error of the order $\sim\mathcal{O}(n^{-2})$ \cite{livro:calculo-numerico},
where $n$ is the number of bins.
We checked for systematic deviation by comparing pairs of $\tau$ estimates
obtained from the same simulation considering a time unit of 1 MCS and 10 MCS.
We found no systematic error for $L\geq 15$ and, since $\tau$ is overestimated,
it does not affect the precision of our equilibrium analysis.
%It is important though to exclude the systems with $L<15$ from our dynamic calculations.

When performing Monte Carlo simulations it is customary to work with the
dimensionless coupling constant $K=J/(k_BT)$ and dimensionless temperature
$\T=1/K$. We also define other dimensionless thermodynamic quantities such as
the extensive energy
\begin{equation}
	E=-\sum_{\left\langle\im,\jm\right\rangle}
	\zeta_{\im\jm}\sigma_\im\sigma_\jm,
\end{equation}
magnetization per spin
\begin{equation}
	m=\frac{1}{N}\sum_{\im=1}^N\sigma_\im,
\end{equation}
specific heat
\begin{equation}
	c=\frac{K^2}{N}\left[\langle E^2\rangle-\langle E\rangle^2\right],
\end{equation}
magnetic susceptibility per spin
\begin{equation}
	\chi =\frac{N}{K}\left[\langle m^2\rangle-\langle |m|\rangle^2\right],
\end{equation}
the quadratic cumulants
\begin{equation}
	U_{22}=\frac{[\langle m^2\rangle^2]-[\langle m^2\rangle]^2}
		{[\langle m^2\rangle]^2}
\end{equation}
and
\begin{equation}
	U_4=\frac{[\langle m^4\rangle]}{[\langle m^2\rangle]^2},
\end{equation}
and the more traditional Binder's cumulant
\begin{equation}\label{eq:binder}
 U\equiv 1-\frac{1}{3}U_4.
\end{equation}
On all equations above, $\langle\cdots\rangle$ denotes thermal averages
whereas $[\cdots]$ denotes averages over disorder configurations.

For each pair of $x$ and $L$ values, we ran simulations at different
temperatures near the critical point (up to 20 temperatures for smaller
lattices and 5 for $L>30$) and used the multiple-histogram
method \cite{artigo:ferrenberg:histograma1,artigo:ferrenberg:histograma2, livro:barkema}
to compute quantities of interest over an almost continuous range of temperatures.
The thermal error associated with those quantities is estimated by dividing the
data from each simulation in blocks and repeating the multiple-histogram process
for each block. The errors are the standard deviation of the values obtained
for a given quantity on the different blocks.
For each set of parameters ($L$, $x$, $T$) we average over $N_S$ samples of
atomic disorder. We chose $10\leq N_S\leq 20$, such that the error due to
disorder was of the same magnitude or smaller than the thermal error obtained
for each disorder configuration. Finally, we sum both thermal and disorder
errors for an estimate of the total error.

\section{Data analysis}\label{data_analysis}
In MC simulations we necessarily deal with finite systems.
However, our interest lies in critical phenomena, which happen in the
thermodynamic limit. The critical behavior of such systems may be extracted from
finite systems by examining the size dependence of the singular part of the free
energy density \cite{livro:julia}.
In this finite-size scaling approach we write the free energy density for a
system of linear size $L$ near the critical point as 
\begin{equation}\label{eq:RGf2}
	\bar f_{\ms{sing}}(t,h,L)
	\sim L^{-d}f^0(tL^{y_t},hL^{y_h},\{ \bar u_iL^{-\omega_i}\})
\end{equation}
where $t$ is the reduced temperature and is given by $(T-T_c)/T_c$,
$H$ is the external magnetic field, and
$h=H/(k_BT)$. We assume that $t$ and $h$
are the only relevant fields while $\bar u_i$ are irrelevant perturbations, such that
$\omega_i>0$, which ensures that in the thermodynamic limit, as
$\bar u_iL^{-\omega_i}\rightarrow 0$, our free energy is a function only of
relevant fields.

Taking appropriate derivatives of the free energy, it is possible to show that
some thermodynamic quantities $Q$ (such as magnetization, specific heat, and magnetic
susceptibility) may be written in the following scaling form:
\begin{equation}\label{eq:FSS_Q}
	Q =
	Q_0L^{\theta} \left\lbrace
	1+\sum_{\im}Q_\im L^{-\omega_\im}
	+\cdots\right\rbrace,
\end{equation}
where $\theta$ is related to the traditional exponents. As (\ref{eq:FSS_Q}) will
be used to fit numerical data to estimate critical exponents, we have to
truncate the sum at some point. As each additional exponent $\omega_i$ taken into
account will add two free parameters to an eventual fit and that reduces
drastically the number of degrees of freedom, we considered only the first
exponent $\omega_1\equiv\omega$. We then have
\begin{align}
	m &= m_0L^{-\beta/\nu}
	\left\lbrace 1+m_1 L^{-\omega}\right\rbrace,
	\label{FSS_mag}
\\
	\chi &= \chi_0L^{\gamma/\nu}
	\left\lbrace 1+\chi_1 L^{-\omega}\right\rbrace,
	\label{FSS_chi}
\\
	c &= c_0L^{\alpha/\nu}
	\left\lbrace 1+c_1 L^{-\omega}\right\rbrace.
	\label{FSS_c}
\end{align}

A similar scaling behavior near the critical point holds for the derivatives
\begin{equation}
	\frac{dG}{dK} = G_0L^{1/\nu}
	\left\lbrace 1+G_1 L^{-\omega}\right\rbrace,
	\label{FSS_dGdK}
\end{equation}
where $G$ stands for the Binder's cumulant $U$ or quantities linked to
the magnetization, such as $\ln{\langle |m|\rangle}$,
$\ln{\langle m^2\rangle}$, and $\ln{\langle |m|^n\rangle}$ \cite{artigo:landau}.
Equation (\ref{FSS_dGdK}) is particularly useful to determine the exponent $\nu$.
Finally, for the critical temperature we have
\begin{equation}
	T_c(L) = T_c+A_0L^{-1/\nu}\left\lbrace 1+A_1 L^{-\omega}\right\rbrace,
	\label{FSS_Tc}
\end{equation}
where $T_c(L)$ is the pseudo-critical temperature for a given system size $L$, and $T_c$ is the true critical temperature.

\begin{figure}[h]
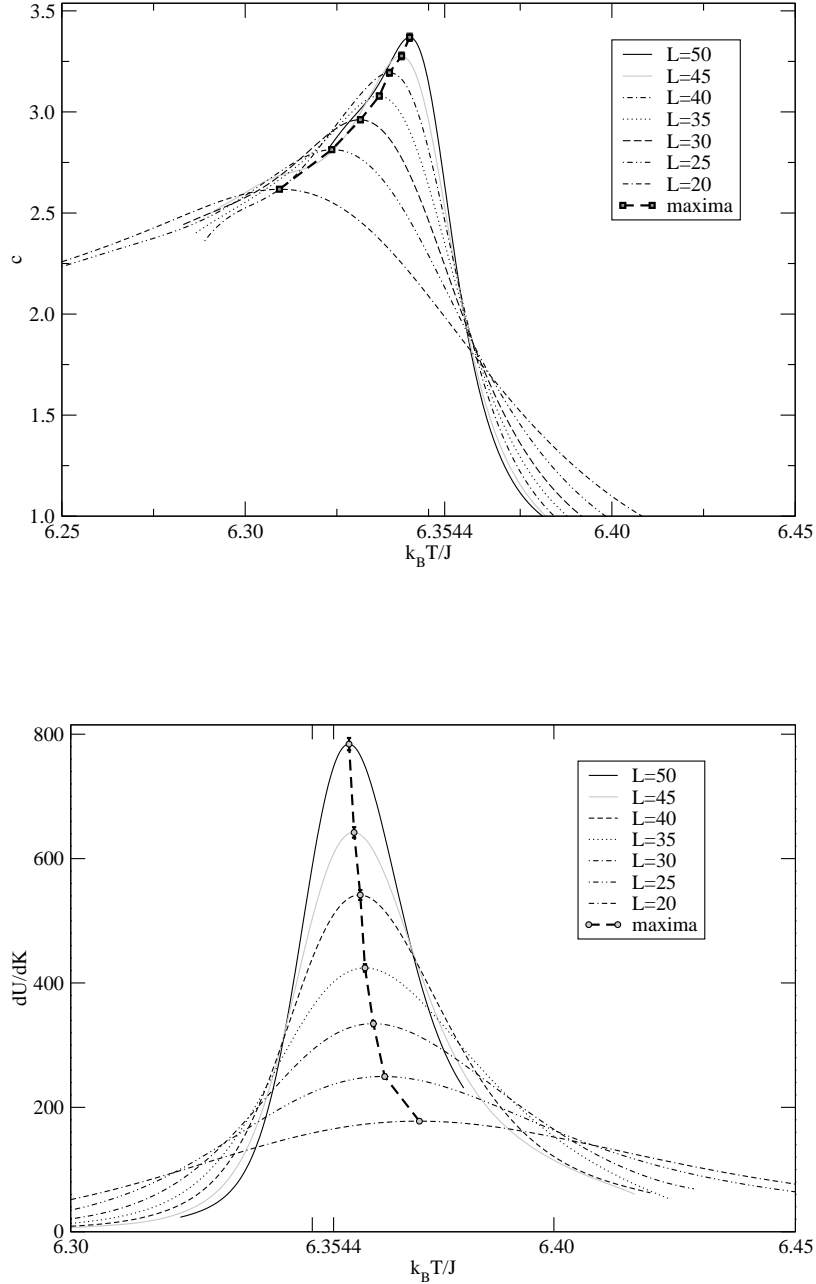

\centering
\includegraphics[width=\figwidth]{fig1a.eps}\\
\vspace{2cm}
\includegraphics[width=\figwidth]{fig1b.eps}
\caption{
Thermodynamic quantities versus dimensionless temperature $\T$ for $x=0\%$ and lattice sizes ranging from 20 to 50.
In (a) we plot the specific heat and in (b) the slope $dU/dK$ of the Binder's cumulant. 
The lines were obtained with the multiple histogram method, the symbols represent
the location of the peaks for each lattice size and the dashed line connecting the peaks is a guide to the eye.
The tick in $\T=6.3544$ corresponds to our estimate of the critical temperature.
The data in this figure and in all other figures in this work were plotted with their respective error bars; 
however, some error bars are smaller then the symbols.}
\label{fig:scale_cal_dbind}
\end{figure}

We perform least square fits using expressions (\ref{FSS_mag})-(\ref{FSS_Tc})
with all quantities computed at several temperatures close to $T_c$.
These estimates of $T_c(L)$ are obtained locating the temperatures $T_m$
where we find the \emph{maxima} of several quantities which diverge as $L\rightarrow\infty$
($c$, $\chi$, $\frac{d\langle m\rangle}{dK}$, $\frac{dU}{dK}$)
as well as the temperatures $T_f$ at which the cumulants $U_4$, $U_{22}$,
and $U_d\equiv U_4-U_{22}$ assume particular fixed values.
The fixed values used were MC estimates of the universal values these
cumulants assume at the critical point on randomly site-diluted (RSIM) and
bond-diluted (RBIM) Ising models: $U_4=1.648(3)$, $U_{22}=0.148(1)$ and
$U_d=1.500(1)$ \cite{hasenbusch2007uc}.
The calculation of the relevant quantities for different temperatures is done using the multiple-histogram method.
Examples of this procedure are shown in Fig. \ref{fig:scale_cal_dbind}
for the specific heat and for the derivative of the Binder cumulant,
both used to locate different estimates of $T_m$.
The comparison between the results from both $T_m$ and $T_f$ FSS methods
provides an additional way to check if our model belongs to the RSIM-RBIM
universality class.

Equations (\ref{FSS_mag})-(\ref{FSS_Tc})
all have four free parameters to be adjusted in the fitting process and no stable fits were possible
for our data, meaning our statistics should be increased if we desire to obtain
the exponents $\alpha$, $\beta$, $\gamma$, $\nu$, and $\omega$  independently. Since we are more interested in 
obtaining $\alpha$, $\beta$, $\gamma$, and $\nu$ than in finding precise correction-to-scaling exponents $\omega$,
we employ a procedure similar to the one presented in Ref. \onlinecite{artigo:landau},
in which we set a fixed value for exponent $\omega$ and perform a fit with three free parameters, instead of four.
Than we change the fixed value of $\omega$ and keep performing fits to obtain the values of the other exponents.
Once several fits are made, we locate the interval of values of $\omega$ such that we minimize the values
of the variance of residuals, $\chi^2/\mbox{DOF}$, where DOF is the number of degrees of freedom of the fit.
We repeat this procedure using system sizes $L_{\smin}\leq L\leq 50$, with $L_{\smin}=$ 5, 10, 12, 15, 18, 20, and 25
to obtain the $L_{\smin}$ that globally minimizes $\chi^2/\mbox{DOF}$.
Once the best $L_{\smin}$ and corresponding series of values for $\omega$ are located, we average the values obtained for the exponents,
making sure to include only fits with $\chi^2/\mbox{DOF}<1.0$ in this statistical analysis.

\section{Results and Discussion}\label{results}

\subsection{Simulation strategy and preliminary results}
For the pure case ($x=0\%$), we ran simulations at $\T=\T_c^{\ms{HTS}}=6.35435$,
which corresponds to the high-temperature series estimate $K_c^{\ms{HTS}}=0.1573725(6)$
of the critical coupling for the pure Ising model on a bcc lattice \cite{artigo:butera}.
We used the single-histogram method \cite{artigo:ferrenberg:histograma1, livro:barkema}
to locate the temperatures where the peaks of the thermodynamic quantities of interest occurred.
From the peak locations we chose temperatures to perform new simulations.
Finally, we employ the multiple-histogram re-weighting
using the data from at least five simulations at different temperatures
such that all peaks were found within the interval between the minimum and maximum of those temperatures.

\begin{figure}[h]
\centering\includegraphics[width=\figwidth]{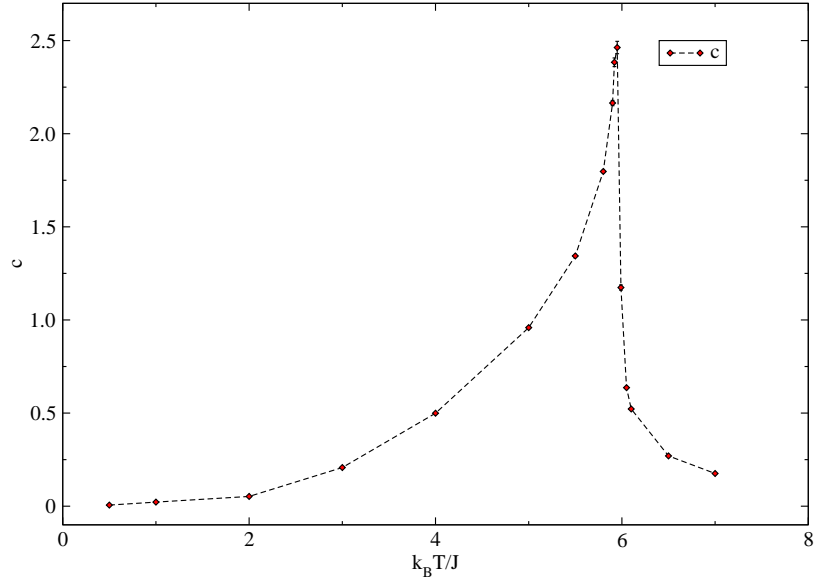}
\caption{Specific heat versus $\T$ for $x=6\%$ and $L=30$.
The points represent the simulation data and the dashed line is a guide to the eye.}
\label{fig:cal_T}
\end{figure}

\begin{figure}[h]
\centering\includegraphics[width=\figwidth]{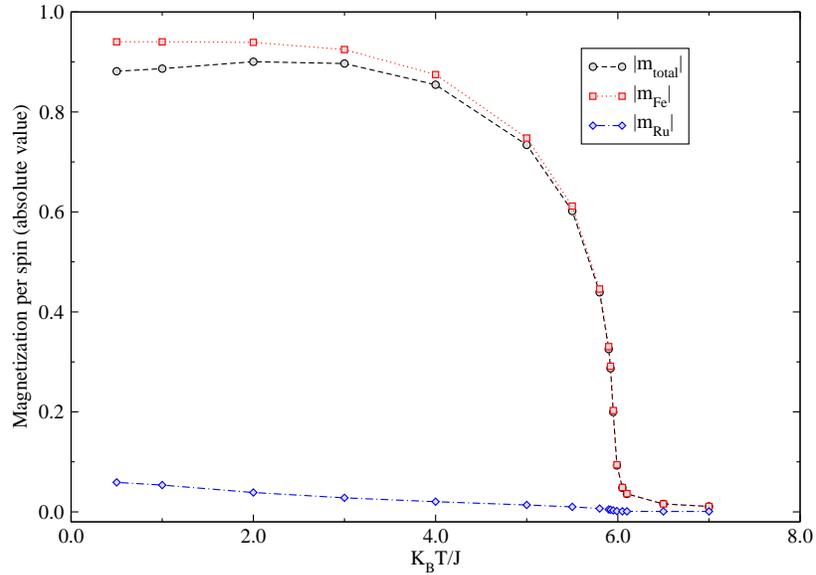}
\caption{(Color online) Total magnetization per spin and individual contributions of \Fe{} and \Ru{}
atoms for the magnetization versus $\T$, obtained for $x=6\%$ and $L=30$.
The points represent the simulation data and the lines are only a guide to the eye.} 
\label{fig:mag_T_MC}
\end{figure}

A similar procedure was employed for the disordered case ($x=4\%$, $6\%$ and $8\%$).
However, since no previous estimates for $T_c$ were available, we performed test simulations over
a wide range of temperatures, in order to make a rough estimate of the location of the critical region.
Figures \ref{fig:cal_T} and \ref{fig:mag_T_MC} exemplify this initial attempt to
determine $T_c(L)$ for $x=6\%$ and $L=30$.
Once determined a suitable temperature interval for each concentration $x$ and size $L$, we divided
it in five ($L>30$) to eleven (smaller lattices) temperatures, to simulate and employ the multiple-histogram 
method as discussed above.

One interesting aspect of Fig. \ref{fig:mag_T_MC}, besides the additional $T_c$ estimate it provides,
is the behavior of the magnetization as a function of temperature. We observe a slight decrease in the total magnetization
at low temperatures, when we approach $T=0$. This decrease is also present in the mean-field approximation
to this model (see Fig. 2 in Ref. \onlinecite{nilton:feru}).
Although not as pronounced as the effect presented in Ref. \onlinecite{nilton:feru}, it also occurs in our Monte Carlo approach;
thus, it is not a mean-field-only effect. For comparison between Fig. \ref{fig:mag_T_MC} and Fig. 2 in Ref. \onlinecite{nilton:feru},
it is important to stress that the lowest temperature we simulated is $\T=0.5$,
which corresponds to $T\approx 80~K$ (this value is obtained using $J=14.16~meV$, as discussed in Sec. \ref{Tc}).

It is possible to present a qualitative description of this effect.
The probability that a \Ru{} atom is completely surrounded by \Fe{} atoms as nearest-neighbors is $(1-x)^8$,
which is quite high for low \Ru{} concentrations
($\approx 51\%$ for $x=8\%$, $\approx 61\%$ for $x=6\%$, $\approx 72\%$ for $x=4\%$, $\approx 85\%$ for $x=2\%$ etc).
Next, let us consider a scenario in which the great majority of \Ru{} atoms have no \Ru{} first neighbors.
In a situation like this, there is close to no frustration and for $T=0$
we expect almost all \Fe{} spins to point in one direction while almost all \Ru{} spins point the other way.
This results in a total magnetization per spin smaller than 1 at $T=0$. 
In fact, if we could guarantee that no \Ru{} atom has a \Ru{} neighbor, we would have $\langle |m|\rangle=1-2x$.
As the temperature rises, thermal fluctuations cause spins to reverse.
Now let us compare two situations: an \Fe{} atom surrounded by \Fe{} atoms, such that all spins are aligned,
and a \Ru{} atom surrounded by \Fe{} atoms, such that the \Ru{} spin is in the opposite direction to the \Fe{} spins.
The probability to reverse the \Fe{} spin is proportional to $\exp{[-8J/k_BT]}$ while
the probability to reverse the \Ru{} spin is proportional to $\exp{[-8\lambda J/k_BT]}$.
As our typical values of $\lambda$ are between $0.0$ and $0.54$, we expect the \Ru{} spins to start reversing
more easily than the \Fe{} spins; thus, the value of $\langle|m|\rangle$ grows with increasing temperature.
However, at higher temperatures the magnetization should decrease, since thermal fluctuations are then strong
enough to flip \Fe{} spins surrounded by \Fe{} atoms.

\subsection{Metropoplis dynamics: choosing spins randomly or sequentially}\label{ran_seq}
In the metropolis algorithm, as originally proposed \cite{artigo:metropolis},
the choice of the atom to be tested for a possible change is usually random.
There are, however, other ways to generate the Markov chain.
It can be done sequentially or, as for the multispin coding version \cite{livro:murilinho},
many spins in a sublattice are tested simultaneously. The above methods are all ergodic and satisfy
detailed balance \cite{livro:barkema} so we ought to choose the most efficient one.

Multispin coding leads to a drastic reduction in computational time for Ising
systems \cite{livro:murilinho} but it is only applicable when it is possible to
express the energy difference between any two given states as an integer,
which is impracticable for our model, since our exchange integrals assume
noninteger values.
We propose an alternative to this method which consists of dividing the lattice
in two sublattices, the same way as multispin coding; however, instead of testing
all spins at once we run over the first sublattice testing all possible spin
flips sequentially and then go to the second sublattice and repeat the procedure.
From now on we will refer to this update scheme as sequential update metropolis,
as opposed to the traditional random update metropolis.

To verify if this sequential method is efficient, compared to the standard
random-update metropolis, we obtain a rough estimate of the dynamical exponent
$z$ for both methods. This is done by fitting our data, at the critical temperature, to the expression
\begin{equation}\label{FSS_tau}
	\tau = AL^z,
\end{equation}
where $A$ is a constant. This is how the correlation time $\tau$ is expected
to behave  at $T_c$, for sufficiently large system sizes $L$.

\begin{figure}[h]
\centering\includegraphics[width=\figwidthgnu,angle=270]{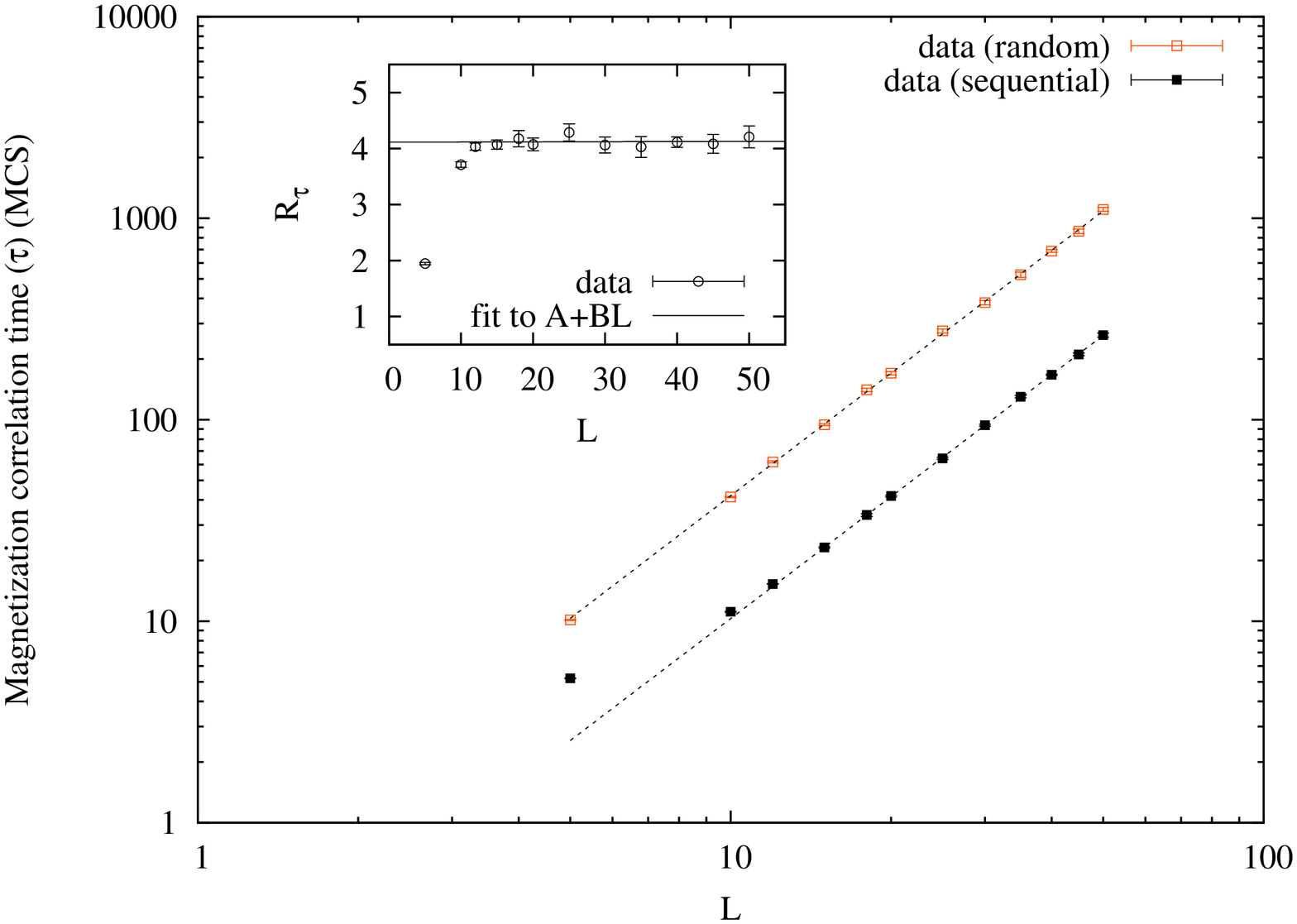}\\
\vspace{1cm}
\includegraphics[width=\figwidthgnu,angle=270]{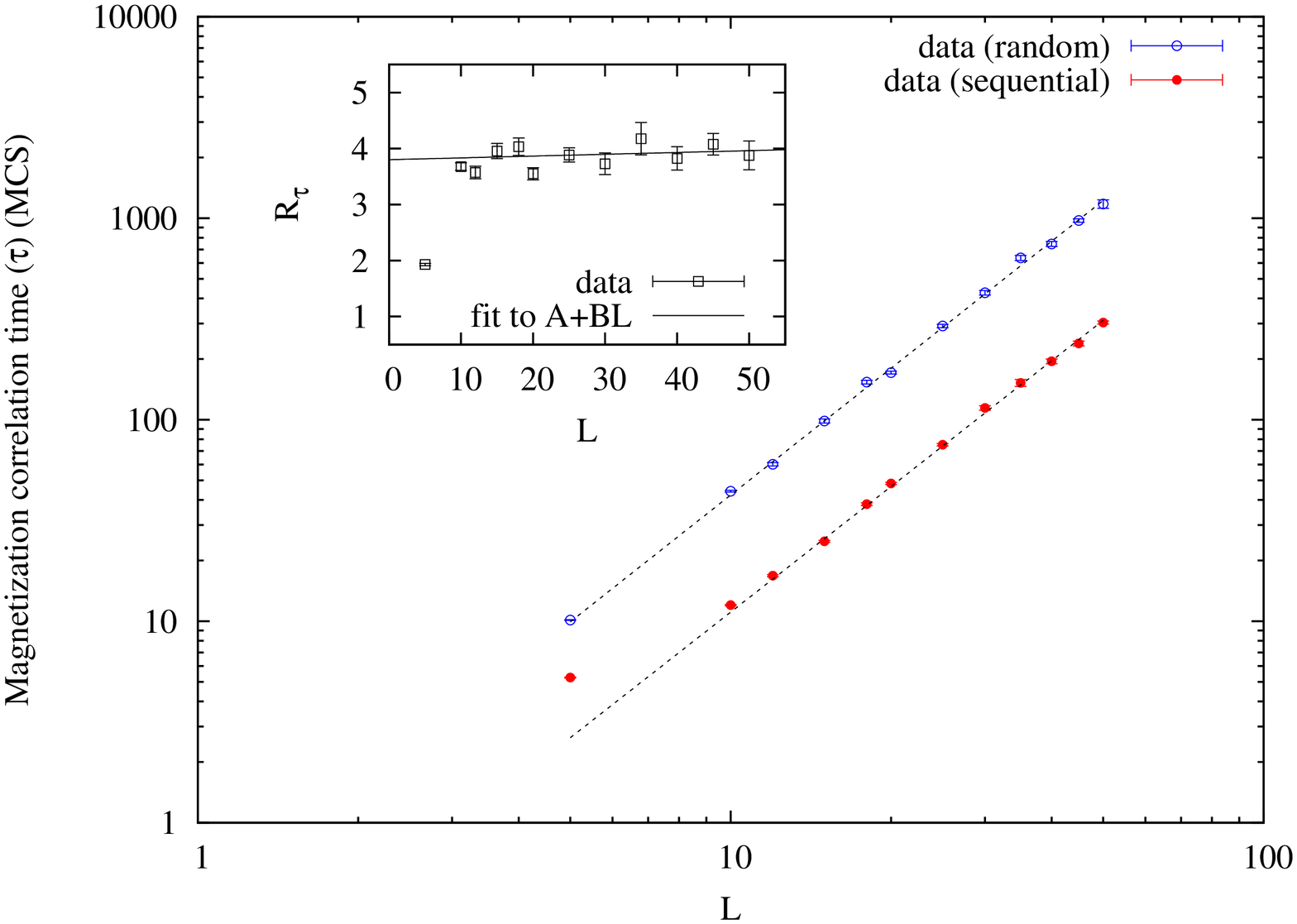}
\caption{(Color online) Log-log plots of the magnetization correlation times versus $L$ for (a) $x=0\%$ and (b) $x=6\%$.
The dashed lines are fits using Eq. (\ref{FSS_tau}).
Each inset is a plot of the ratio between random and sequential correlation times versus $L$
for the respective ruthenium concentration where the solid line is a fit using $R_{\tau}=A+BL$.
All fits were done for $L\geq 15$, to avoid the systematic
error introduced by numerical integration, present for correlation times smaller than or equal to 10 MCS,
as discussed in Sec. \ref{monte_carlo}.}
\label{fig:tau_mag}
\end{figure}

In Fig. \ref{fig:tau_mag}, we present fits for the correlation times.
We note that the curves for both random and sequential correlation times have almost the same slope,
which indicates that random and sequential algorithms have approximately the same
dynamic exponent $z$, as expected. However, random correlation times are always
larger than sequential ones for a fixed system size.
In fact we also plot the ratio
\begin{equation}\label{FSS:tau_ratio}
	R_{\tau}=\frac{\tau_{\ms{ran}}}{\tau_{\ms{seq}}}
\end{equation}
as a function of $L$ and find an almost-constant $R_{\tau}\approx 4$, as
shown in the insets of Figs. \ref{fig:tau_mag} (a) and \ref{fig:tau_mag} (b). For $L \geq 15$, we have
fit the data to $R_{\tau}=A+BL$ and obtained $A=4.11(8)$, $B=0.000(3)$
for the pure case and $A=3.8(2)$, $B=0.003(6)$ for the disordered case. 

The estimates we get for the dynamical exponent for the pure case are
$z=2.02(1)$ for random updates and $z=2.01(1)$ for sequential updates.
For the disordered case we have only estimated $z$ for $x=6\%$, which gave us the
figures $z=2.06(3)$ and $z=2.02(2)$ for the random and sequential update
dynamics, respectively.
All values are in agreement with the MC result, $z=2.04(3)$, for the three-dimensional (3D) pure Ising model,
presented in Ref. \onlinecite{artigo:landau-z}.
So, within error bars, both pure and disordered systems
have the same dynamic exponent for random or sequential updates.
However, we find that the correlation times for a given system size $L$ are always
greater for the random update scheme. Therefore, the sequential update version
is more efficient than the random update version.

\subsection{Critical exponents}\label{exponents}

\begin{figure}[h]
\centering\includegraphics[width=\figwidthgnu,angle=270]{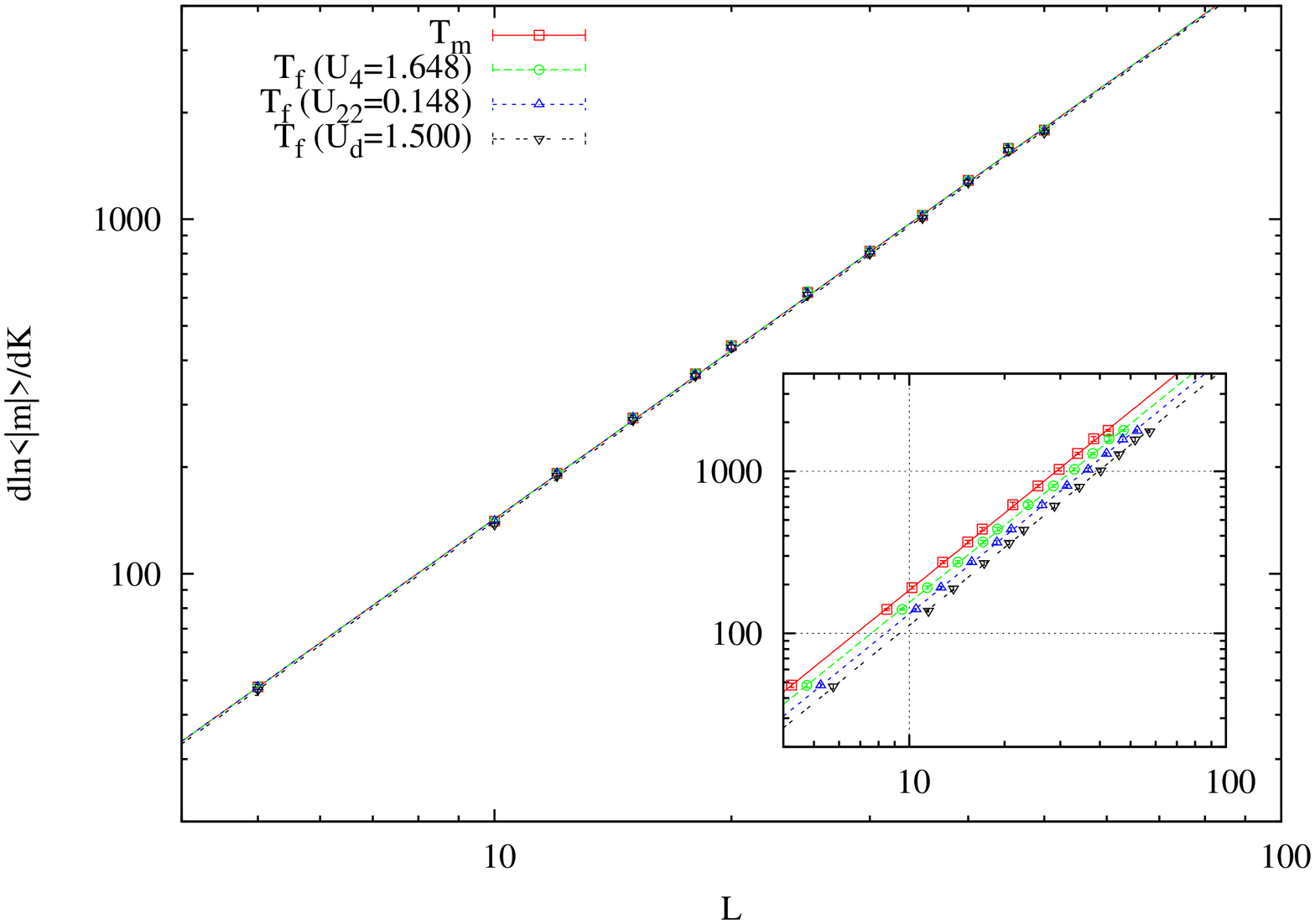}\\
\vspace{1cm}
\includegraphics[width=\figwidthgnu,angle=270]{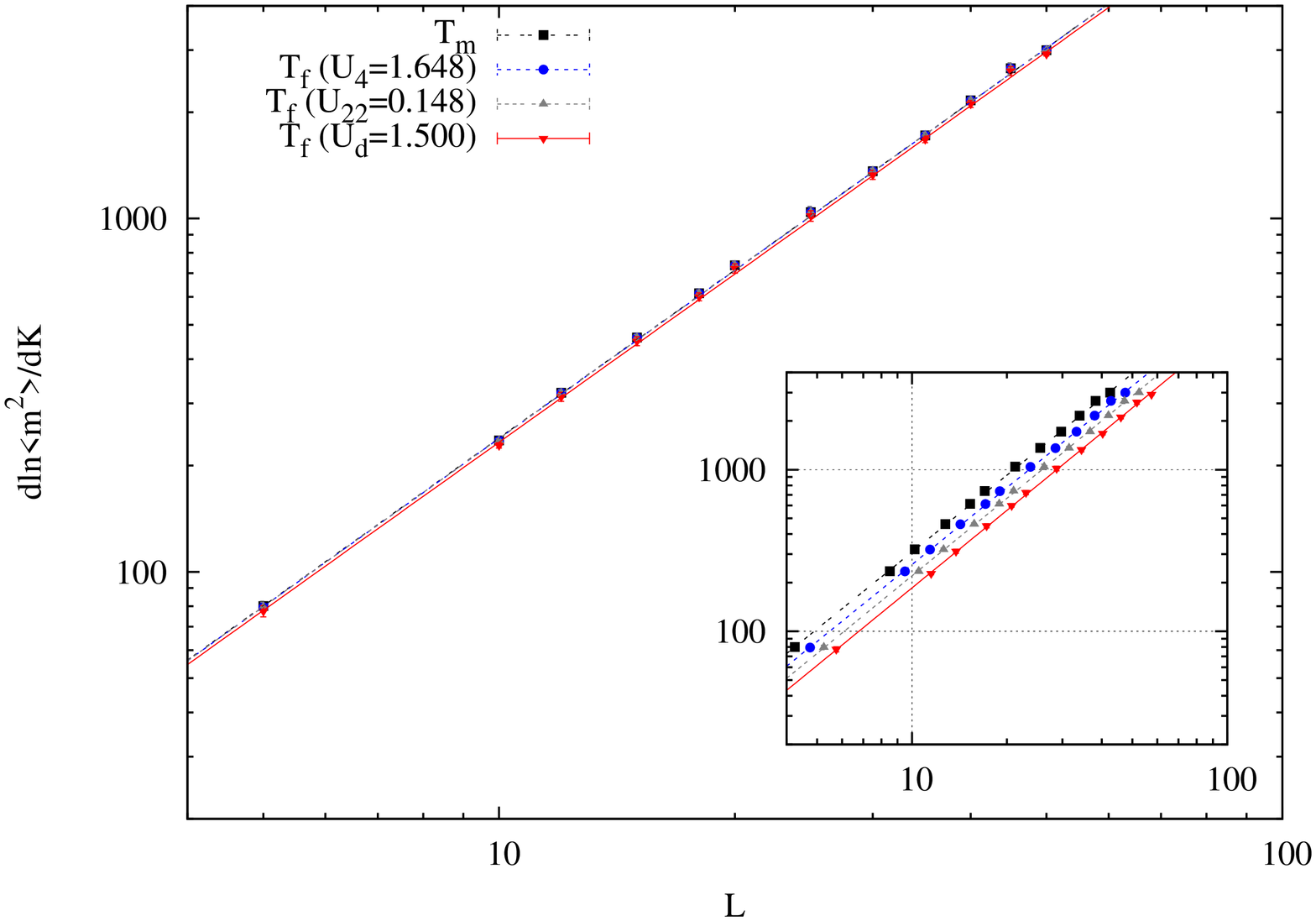}
\caption{(Color online) Log-log plot of the derivative $dG/dK$ versus $L$ for $x=4\%$, for
(a) $G=\ln{\langle|m|\rangle}$ and  (b) $G=\ln{\langle m^2\rangle}$
[see Eq. (\ref{FSS_dGdK}) and remarks after it].
The data were computed at slightly different critical point estimates,
given by the temperatures $T_m$ (maximum of $dG/dK$) and $T_f$, in order to compare the estimates.
The solid lines are fits to $dG/dK=a_0L^{1/\nu}$.
The inset corresponds to the same graph made with both data and lines slightly shifted along the $L$ axis to make them visible.}
\label{fig:Tm_Tf}
\end{figure}

Figure \ref{fig:Tm_Tf} shows the comparison between the behavior of some quantities $dG/dK$
at different estimates of the critical point, obtained by the methods described in Sec. \ref{data_analysis}.
Note how the quantities assume almost the same value for the same size at the four different $T_c$ estimates.
Therefore, the four different fits for each quantity are almost indistinguishable.
As a result, the numerical values we obtain for each critical exponent by independent methods
are the same, within error bars. Thus, we combine both $T_m$ and $T_f$ FSS methods to obtain our final estimates
of $\alpha$, $\gamma$, $\beta$, and $\nu$ for the disordered systems.

%%%%%%%%%%%%%%%%%%%%%%%%%%%%%%%%%%%
%% TAB
\begin{table}[h]\caption{Estimates of $1/\nu$ using Eq. (\ref{FSS_dGdK}) for some quantities $G$, with $L_{\smin}=15$.
The region $0.30 \leq\omega\leq 0.36$ minimized the values of $\chi^2/\mbox{DOF}$.
In this interval we found no difference between $\chi^2/\mbox{DOF}$ up to the second decimal figure
for each quantity considered. The values obtained were $\chi^2/\mbox{DOF}=0.22$, $0.23$, and $0.081$
for $G=\ln{\langle|m|\rangle}$, $G=\ln{\langle m^2\rangle}$, and $G=U$, respectively. 
}
\begin{tabular}{cccc}
	\hline\hline
	& \multicolumn{3}{c}{$1/\nu$}\\
	\hline
	$\omega$ & $G=\ln{\langle|m|\rangle}$ & $G=\ln{\langle m^2\rangle}$ & $G=U$\\
	\hline
	0.300 & 1.465(31) & 1.463(29) & 1.444(47)\\
	0.305 & 1.466(31) & 1.464(29)  & 1.445(46)\\
	0.310 & 1.467(31) & 1.465(29) & 1.446(46)\\
	0.315 & 1.468(31) & 1.466(29) & 1.448(46)\\
	0.320 & 1.469(31) & 1.467(28) & 1.448(46)\\
	0.325 & 1.470(30) & 1.468(28) & 1.449(45)\\
	0.330 & 1.471(30) & 1.469(28) & 1.450(45)\\
	0.335 & 1.472(30)  & 1.470(28) & 1.451(45)\\
	0.340 & 1.473(30) & 1.471(28) & 1.452(45)\\
	0.345 & 1.474(30) & 1.472(28) & 1.453(44)\\
	0.350 & 1.474(30) & 1.473(27) & 1.454(44)\\
	0.355 & 1.475(29) & 1.474(27) & 1.455(44)\\
	0.360 & 1.476(29) & 1.474(27) & 1.456(44)\\
	\hline\hline\\
\end{tabular}
\label{tab:x0.06_fit-nu-omega}
\end{table}
%%%%%%%%%%%%%%%%%%%%%%%%%%%%%%%%%%

\begin{figure}[h]
\centering\includegraphics[width=\figwidth]{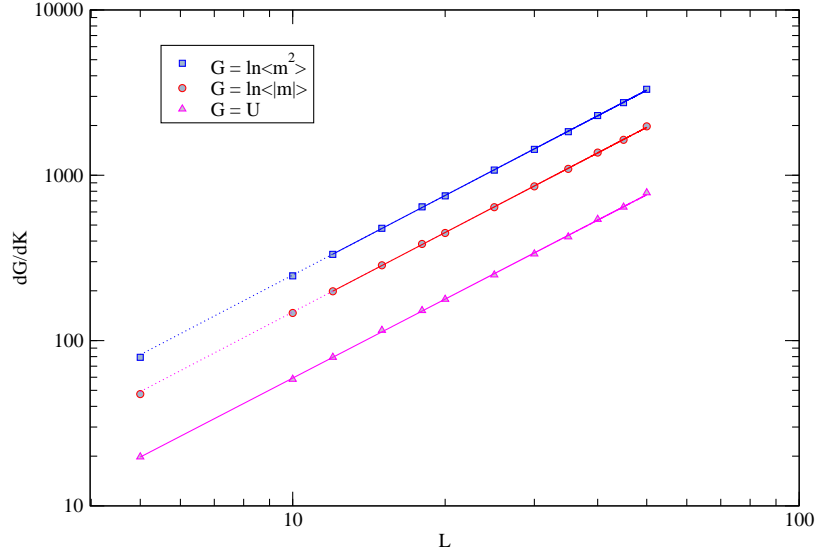}\\
\vspace{2cm}
\includegraphics[width=\figwidth]{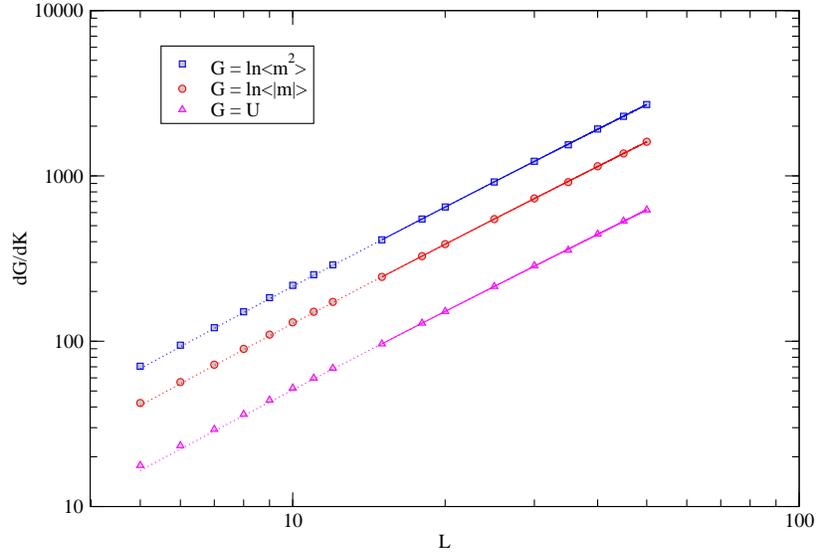}
\caption{(Color online) Log-log plots of the \emph{maxima} of some thermodynamic quantities $G$ as functions of $L$
that were used to determine $\nu$ for (a) $x=0\%$ and (b) $x=6\%$.
The full lines are fits performed with Eq. (\ref{FSS_dGdK}) for $L_{\smin}<L<50$.
The dotted lines are extrapolations of the fits for $L<L_{\smin}$.}
\label{fig:fit-nu-omega}
\end{figure}

To estimate $\nu$ we fit the data to Eq. (\ref{FSS_dGdK}), where we used
$G=U$, $G=\ln{\langle |m|\rangle}$, and $G=\ln{\langle m^2\rangle}$.
For $x=0\%$ we computed the derivatives at the temperature $T_m$ where we found the maximum of each quantity.
Following the procedure described in Sec. \ref{data_analysis}, we found good fits
with $L_{\smin}=12$ for the logarithmic derivatives and $L_{\smin}=5$ for the Binder's cumulant.
The fits in Fig. \ref{fig:fit-nu-omega} (a)
were done with $\omega\approx 1.0$, consistent with the result reported in Ref. \onlinecite{artigo:landau}.  
The value we obtain in this work, $\nu=0.6269(20)$, is in excellent agreement with others in the literature
for the three-dimensional Ising model \cite{artigo:landau,
compostrini1999,artigo:butera,guillou-zinn1980,guillou-zinn1987,belanger2000}.

For the disordered cases, fits using Eq. (\ref{FSS_dGdK}) were done with the same quantities
$G=U$, $G=\ln{\langle |m|\rangle}$, and $G=\ln{\langle m^2\rangle}$.
Tab. \ref{tab:x0.06_fit-nu-omega} shows our results for $1/\nu$ for $x=6$ $\%$, obtained from data at
temperatures $T_m$.
One of these fits is shown in Fig. \ref{fig:fit-nu-omega} (b).
Combining the various estimates in Tab. \ref{tab:x0.06_fit-nu-omega}, we find $1/\nu=1.4632(90)$ or $\nu=0.6834(42)$.
Using both $T_m$ and $T_f$ methods, we arrive at our final estimate, $\nu=0.6826(46)$,
presented in Tab. \ref{tab:exponents} along with our final estimates of $\nu$ for the other \Ru{} concentrations.

%%%%%%%%%%%%%%%%%%%%%%%%%%%
%% TAB
\begin{table}[h]
\caption{Our estimates of the exponents $\alpha$, $\beta$, $\gamma$, and $\nu$ compared to other works in the literature.
The lines with the labels $x=0\%$, $4\%$, $6\%$, and $8\%$ correspond to our final estimates.
The table also contains other
Monte Carlo \cite{artigo:landau, compostrini1999},
high-temperature series expansion\cite{artigo:butera}, and
quantum field theory \cite{guillou-zinn1980,guillou-zinn1987}
results for the pure 3D Ising universality class, as well as
Monte Carlo results for the RSIM \cite{ballesteros1998},
RBIM \cite{calabrese2003tdr,hasenbusch2007uc}, and
$\pm J$ Ising Model \cite{hasenbusch2007+-J}. Reference \onlinecite{belanger2000}
reports experimental results for both pure and disordered (with low disorder) 
3D Ising universality classes.}

\begin{tabular*}{0.9\textwidth}{@{\extracolsep{\fill}}ccccccc}
	\hline\hline
	\multicolumn{7}{c}{Pure}\\
	& $\alpha$ & $\beta$ & $\gamma$ & $\nu$ & $\alpha+2\beta+\gamma$ & $2-\alpha-3\nu$\\
	\hline
	$x=0\%$ & 0.1093(47) & 0.3316(86) & 1.2448(70) & 0.6269(20) & 2.017(29) & 0.010(11)\\
	\cite{artigo:landau} & 0.1190(60) & 0.3258(44) & 1.2390(71) & 0.627(2) & 2.020(21) & $\equiv 0$\footnote{The
label $\equiv 0$ corresponds to cases where $\alpha$ was calculated using the Josephson equality.}\\
	\cite{compostrini1999} & 0.1099(7) & 0.32648(18) & 1.2371(4) & 0.63002(23) & $\equiv 2$\footnote{The
label $\equiv 2$ corresponds to cases where either $\beta$ or $\gamma$ was calculated using the Rushbrooke equality.} & $\equiv 0$\\
	\cite{artigo:butera} & 0.1094(45) & 0.3266(10) & 1.2375(6) & 0.6302(4) & $\equiv 2$ & $\equiv 0$\\
	\cite{guillou-zinn1980,guillou-zinn1987} & 0.1100(45) & 0.3270(15) & 1.2390(25) & 0.6300(15) & 2.003(10) & $\equiv 0$\\
	\cite{belanger2000} & 0.110(5) & 0.325(5) & 1.25(2) & 0.64(1) & 2.01(4) & -0.03(4)\\
	\hline
	\multicolumn{7}{c}{Disordered}\\
	& $\alpha$	& $\beta$	& $\gamma$	& $\nu$ & $\alpha+2\beta+\gamma$ & $2-\alpha-3\nu$\\
	\hline
	$x=4\%$	& -0.062(25) & 0.3584(96) & 1.334(20) & 0.6873(84) & 1.989(64) & $\equiv 0$\\
	$x=6\%$	& -0.049(15) & 0.359(16) & 1.330(18) & 0.6826(46) & 1.999(65) & $\equiv 0$\\
	$x=8\%$	& -0.057(20) & 0.3588(81) & 1.337(21) & 0.6856(66) & 2.003(46) & $\equiv 0$\\
	\cite{ballesteros1998}	& -0.051(16) & 0.3546(28) & 1.342(10) & 0.6837(53) & 1.994(32) & 0.006(32) \\
	\cite{calabrese2003tdr}	& -0.049(9) & 0.3535(17) & 1.342(6) & 0.683(3) & $\equiv 2$ & $\equiv 0$\\
	\cite{hasenbusch2007uc}	& -0.049(6) & 0.354(1) & 1.341(4) & 0.683(2) & $\equiv 2$ & $\equiv 0$\\
	\cite{hasenbusch2007+-J}& -0.046(6) & 0.329(2) & 1.339(7) & 0.682(2) & 1.951(17) & $\equiv 0$\\
	\cite{belanger2000}	& -0.10(2) & 0.350(9) & 1.31(3) & 0.69(1) & 1.91(7) & 0.03(5)\\
	\hline\hline
\end{tabular*}
\label{tab:exponents}
\end{table}
%%%%%%%%%%%%%%%%%%%%%%%%%%

\begin{figure}[h]
\centering\includegraphics[width=\figwidth]{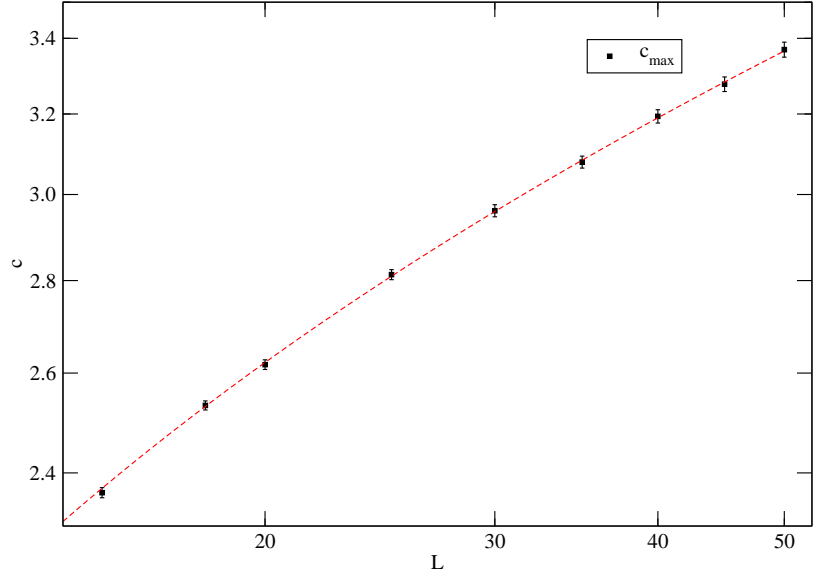}\\
\vspace{2cm}
\includegraphics[width=\figwidth]{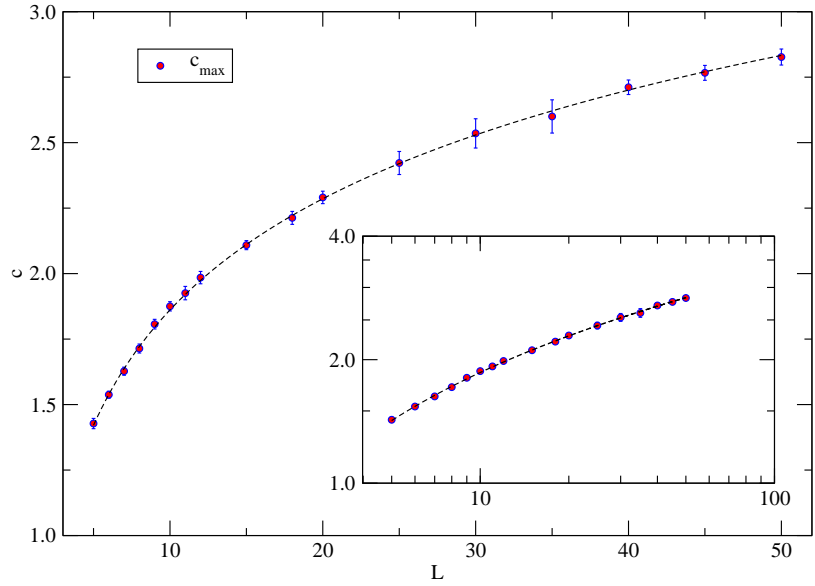}
\caption{(Color online) \emph{Maxima} of the specific heat versus $L$.
For $x=0\%$ (a), the plot is in log-log scale and the dashed line is a fit of the data using
Eq. (\ref{FSS_c}), with $\omega=1.0$.
For $x=6\%$ (b), the plot is on a linear scale and the inset corresponds to the same data on a log-log scale.
The dashed line in (b) is only a guide to the eye.}
\label{fig:fit_cal}
\end{figure}

Following the same procedures, we used Eq. (\ref{FSS_c}) to find $\alpha$.
For $x=0\%$, the results are presented in Tab. \ref{tab:exponents} and compared to other results in the literature. 
Figure \ref{fig:fit_cal} (a)
shows one of the fits for the specific heat of the pure system:
we obtain the estimate $\alpha/\nu=0.1743(70)$. For the disordered systems,
however, no stable fits were possible with only one correction-to-scaling exponent and we lack statistical resolution
to perform fits with higher-order correction-to-scaling terms.
Fig. \ref{fig:fit_cal} (b)
shows one of the plots for the specific heat of the disordered system ($x=6\%$)
in which the dashed line is only a guide to the eye.

The estimates for $\alpha$ for the disordered case, presented in Tab. \ref{tab:exponents},
were not determined independently, but were obtained using the Josephson equality
$\alpha=2-\nu d$, where $d$ is the dimension ($d=3$ in our case).
The $\beta$ and $\gamma$ values were all obtained independently by fitting Eqs. (\ref{FSS_mag}) and (\ref{FSS_chi}),
respectively, and our final estimates are presented in Tab. \ref{tab:exponents}.
We see that, within error bars, critical exponents do not depend on the concentration $x$, as expected.
Note that the usual scaling relations for the exponents are satisfied, for both pure and disordered cases.
This result is an independent check of our calculation. Also, our results for the critical exponents agree, within
error bars, with values reported previously elsewhere,
obtained both from theoretical methods or from experimental studies.
This gives support to our hypothesis concerning the universality class
of the model treated in this work.

\subsection{Critical temperatures}\label{Tc}

\begin{figure}[h]
\centering\includegraphics[width=\figwidth]{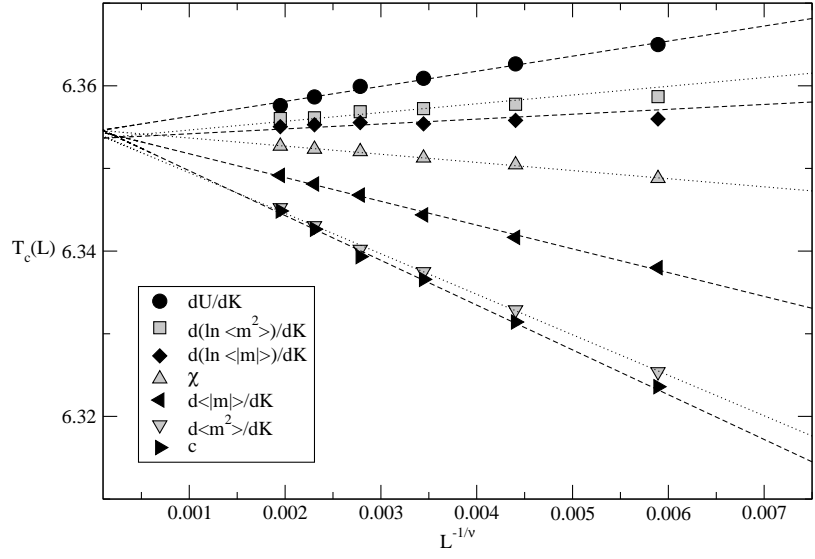}
\caption{Effective critical temperatures $T_c(L)$ versus $L^{-1/\nu}$ estimated for several quantities.
The dotted and dashed lines are fits performed with Eq. \ref{FSS_Tc} and using our estimate $\nu=0.6269$.}
\label{fig:x0.0fittc}
\end{figure}

For $x=0\%$, we fit Eq. (\ref{FSS_Tc}) for the temperatures where we located peaks of
several thermodynamic quantities. Figure \ref{fig:x0.0fittc} shows the values of $T_m$ used in our FSS analysis.
Our best fits were obtained with $L_{\smin}=25$ for the magnetic susceptibility and the derivatives of
$\langle|m|\rangle$ and $U$ and $L_{\smin}=20$ for the remaining quantities.
Combining the estimates obtained in those fits, presented in Tab. \ref{tab:x0.0_fit-tc},
we obtain: $\T_c=6.3544(6)$, or $K_c=0.15737(1)$, which is in excellent agreement with
$K_c^{\ms{HTS}}=0.1573725(6)$ \cite{artigo:butera} and with another Monte Carlo evaluation
\cite{lundow2009}, $K_c=0.157371(1)$.
We can use our $\T_c$ to estimate the exchange integral of the \Fe-\Fe{} bond, $J_{\ms{FeFe}}\equiv J$,
just by combining the experimental $T_c$ in Tab. \ref{tab:Tc(x)} with the definition $\T=k_BT/J$.
We obtain $J=14.16$ meV, which is close to $12.9$ meV, reported in Ref. \onlinecite{nilton:feru}
and in the interval 10 to 50 meV, as expected for \Fe{}, \Co{}, and \Ni{} \cite{kaul1983}.

%%%%%%%%%%%%%%%%%%%%%%%%%%%%
%% TAB
\begin{table}[h]
\caption{Estimates of $T_c(x=0\%)$ obtained with FSS analysis of $T_c(L)$ for some thermodynamic quantities.}
\begin{tabular}{lcc}
	\hline\hline
	Quantity & $L_{\smin}$ & $\T_c$\\
	\hline
		$c$					& 20 &	6.35514(34)\\
		$\chi$					& 25 &	6.35467(11)\\
		$\frac{dU}{dK}$				& 25 &	6.35447(37)\\
		$\frac{d\langle|m|\rangle}{dK}$		& 25 &	6.35465(34)\\
		$\frac{d}{dK}\ln{\langle|m|\rangle}$	& 20 &	6.35360(57)\\
		$\frac{d\langle m^2\rangle}{dK}$	& 20 &	6.35437(26)\\
		$\frac{d}{dK}\ln{\langle m^2\rangle}$	& 20 &	6.35359(58)\\
	\hline\hline
\end{tabular}
\label{tab:x0.0_fit-tc}
\end{table}
%%%%%%%%%%%%%%%%%%%%%%%%%%%%

%%%%%%%%%%%%%%%%%%%%%%%%%
%% TAB
\begin{table}[h]
\caption{Our Monte Carlo estimates of critical temperatures compared to experimental and mean-field results.}
\begin{tabular}{ccccc}
\hline\hline
  & \multicolumn{2}{c}{This work} & Experimental \cite{artigo:paduani} & Mean-field \cite{nilton:feru}\\
$x$ & $\T_c$ & $T_c (K)$ & $T_c (K)$ & $T_c (K)$\\
\hline
$0\%$  & 6.3544(6)  & -         & 1043   & 1043\\
$2\%$  & -          & -         & 968(2) & 983\\
$4\%$  & 6.1089(23) & 1002.7(4) & 928(2) & 933\\
$6\%$  & 5.9615(17) & 978.5(3)  & 908(2) & 893\\
$8\%$  & 5.8064(22) & 953.0(4)  & -      & 863\\
$10\%$ & -          & -         & 838(2) & 842\\
\hline\hline
\end{tabular}
\label{tab:Tc(x)_MC}
\end{table}
%%%%%%%%%%%%%%%%%%%%%%%%%

Following the same procedure for the disordered systems and also including the temperatures $T_f$ (fixed values of the cumulants) in the analysis,
we find the critical temperatures for $x=4\%$, $6\%$, and $8\%$.
The value $J=14.16$ meV is used to calculate all critical temperatures in Kelvin, in order to compare with experimental results,
as presented in Tab. \ref{tab:Tc(x)_MC}. The mean-field critical temperatures are easily computed with Eq. (\ref{eq:fit}),
which is taken from Ref. \onlinecite{nilton:feru} [see Eq. (14) in Ref. \onlinecite{nilton:feru}].

Our Monte Carlo estimates of $T_c$ for the disordered cases
do not agree with the mean-field prediction; neither do they agree with the experimental values.
We ascribe this discrepancy to the method by which the parameters of the model were obtained.
Although those parameters fit the experimental data well in the mean-field approach,
the model is not suitable for these \feru{} alloys in the context of Monte Carlo simulations.

\section{Conclusion}\label{conclusion}

In this study we used Monte Carlo simulations to investigate the magnetic properties and critical behavior
of a model proposed by Paduani and Branco (2008) for \feru{} alloys through a mean-field approach.
Our simulations were restricted to ruthenium concentrations $x=0\%$, $4\%$, $6\%$, and $8\%$.
We employed re-weighting single and multiple histogram methods and finite-size scaling analysis,
considering up to the first-order correction-to-scaling exponent in order to obtain the critical
temperature and critical exponents of the model. 

In the pure case, $x=0\%$, the values ​​obtained for the critical exponents are in excellent agreement
with the ones in the literature.
The critical temperature we found for the pure system also agrees very well with the
high-temperature series expansion estimate by Butera and Comi (2000) and with another Monte Carlo result \cite{lundow2009}.
In the disordered cases, we show that the critical exponents are consistent with the universality class of the transition
line between paramagnetic and ferromagnetic phases of three-dimensional Ising models with random site or random bond dilution,
as well as with the three-dimensional Ising model with randomly distributed $+J$ and $-J$ exchange integrals.

For $x=4\%$, $6\%$, and $8\%$, our estimates of $T_c$ do not agree with the mean-field prediction 
nor with experimental results.
This is expected, since the model parameters were determined through a fitting procedure of experimental
data in a mean-field approach \cite{nilton:feru}.
However, the model proposed in Ref. \onlinecite{nilton:feru} is in the universality class
of three-dimensional disordered Ising models. Furthermore, our values for the critical exponents do not depend on $x$,
as expected. This result is obtained only if we take into account a correction-to-scaling term in the finite-size-scaling
analysis. 

To propose a model that brings simulations and experimental results closer together,
we must seek other ways to determine the dependence of the exchange integrals of \Fe-\Ru{} and \Ru-\Ru{} bonds
with the ruthenium concentration. One possibility would be a mean-field renormalization group approach,
which is presently being carried out.

\begin{acknowledgments}
This work has been partially supported by Brazilian Agencies FAPESC, CNPq, and CAPES.
The authors would also like to thank
A. Caparica, W. Figueiredo, C. Paduani and D. Girardi
for assistance with the manuscript.
\end{acknowledgments}

%Merlin.mbs v4.21 2009-07-09.
%

%%%%%%%%%%
%%% FIGS.
%%%%%%%%%%
\newpage

%%%%%%%%%%%%%%%%%%%%%%%%%%%%%%%
%%% FIG 1
%%%%%%%%%%%%%%%%%%%%%%%%%%%%%%%
%%%%%%%%%%%%%%%%%%%%%%%%%%%%%%%%

%%%%%%%%%%%%%%%%%%%%%%%%%%%%%%%
%%% FIG 2
%%%%%%%%%%%%%%%%%%%%%%%%%%%%%%%
%%%%%%%%%%%%%%%%%%%%%%%%%%%%%%%

%%%%%%%%%%%%%%%%%%%%%%%%%%%%%%%
%%% FIG 3
%%%%%%%%%%%%%%%%%%%%%%%%%%%%%%%
%%%%%%%%%%%%%%%%%%%%%%%%%%%%%%%

%%%%%%%%%%%%%%%%%%%%%%%%%%%%%%%
%%% FIG 4
%%%%%%%%%%%%%%%%%%%%%%%%%%%%%%%
%%%%%%%%%%%%%%%%%%%%%%%%%%%%%%%

%%%%%%%%%%%%%%%%%%%%%%%%%%%%%%%
%%% FIG 5
%%%%%%%%%%%%%%%%%%%%%%%%%%%%%%%
%%%%%%%%%%%%%%%%%%%%%%%%%%%%%%%%

%%%%%%%%%%%%%%%%%%%%%%%%%%%%%%%
%%% FIG 6
%%%%%%%%%%%%%%%%%%%%%%%%%%%%%%%
%%%%%%%%%%%%%%%%%%%%%%%%%%%%%%%

%%%%%%%%%%%%%%%%%%%%%%%%%%%%%%%
%%% FIG 7
%%%%%%%%%%%%%%%%%%%%%%%%%%%%%%%
%%%%%%%%%%%%%%%%%%%%%%%%%%%%%%%

%%%%%%%%%%%%%%%%%%%%%%%%%%%%%%%
%%% FIG 8
%%%%%%%%%%%%%%%%%%%%%%%%%%%%%%%
%%%%%%%%%%%%%%%%%%%%%%%%%%%%%%%


\begin{thebibliography}{10}%
\makeatletter
\providecommand \@ifxundefined [1]{%
 \ifx #1\undefined \expandafter \@firstoftwo
 \else \expandafter \@secondoftwo
\fi
}%
\providecommand \@ifnum [1]{%
 \ifnum #1\expandafter \@firstoftwo
 \else \expandafter \@secondoftwo
\fi
}%
\providecommand \enquote [1]{``#1''}%
\providecommand \bibnamefont  [1]{#1}%
\providecommand \bibfnamefont [1]{#1}%
\providecommand \citenamefont [1]{#1}%
\providecommand\href[0]{\@sanitize\@href}%
\providecommand\@href[1]{\endgroup\@@startlink{#1}\endgroup\@@href}%
\providecommand\@@href[1]{#1\@@endlink}%
\providecommand \@sanitize [0]{\begingroup\catcode`\&12\catcode`\#12\relax}%
\@ifxundefined \pdfoutput {\@firstoftwo}{%
 \@ifnum{\z@=\pdfoutput}{\@firstoftwo}{\@secondoftwo}%
}{%
 \providecommand\@@startlink[1]{\leavevmode}%
 \providecommand\@@endlink[0]{}%
}{%
 \providecommand\@@startlink[1]{%
  \leavevmode
  \pdfstartlink
   attr{/Border[0 0 1 ]/H/I/C[0 1 1]}%
   user{/Subtype/Link/A<</Type/Action/S/URI/URI(#1)>>}%
  \relax
 }%
 \providecommand\@@endlink[0]{\pdfendlink}%
}%
\providecommand \url  [0]{\begingroup\@sanitize \@url }%
\providecommand \@url [1]{\endgroup\@href {#1}{\urlprefix}}%
\providecommand \urlprefix [0]{URL }%
\providecommand \Eprint[0]{\href }%
\@ifxundefined \urlstyle {%
  \providecommand \doi [1]{doi:\discretionary{}{}{}#1}%
}{%
  \providecommand \doi [0]{doi:\discretionary{}{}{}\begingroup
  \urlstyle{rm}\Url }%
}%
\providecommand \doibase [0]{http://dx.doi.org/}%
\providecommand \Doi[1]{\href{\doibase#1}}%
\providecommand \bibAnnote [3]{%
  \BibitemShut{#1}%
  \begin{quotation}\noindent
    \textsc{Key:}\ #2\\\textsc{Annotation:}\ #3%
  \end{quotation}%
}%
\providecommand \bibAnnoteFile [2]{%
  \IfFileExists{#2}{\bibAnnote {#1} {#2} {\input{#2}}}{}%
}%
\providecommand \typeout [0]{\immediate \write \m@ne }%
\providecommand \selectlanguage [0]{\@gobble}%
\providecommand \bibinfo [0]{\@secondoftwo}%
\providecommand \bibfield [0]{\@secondoftwo}%
\providecommand \translation [1]{[#1]}%
\providecommand \BibitemOpen[0]{}%
\providecommand \bibitemStop [0]{}%
\providecommand \bibitemNoStop [0]{.\EOS\space}%
\providecommand \EOS [0]{\spacefactor3000\relax}%
\providecommand \BibitemShut [1]{\csname bibitem#1\endcsname}%
%</preamble>
\bibitem{nilton:feru}%
  \BibitemOpen
  \bibfield{author}{%
  \bibinfo {author} {\bibfnamefont{C.}~\bibnamefont{Paduani}}\ and\ \bibinfo
  {author} {\bibfnamefont{N.~S.}\ \bibnamefont{Branco}},\ }%
  \bibfield{journal}{%
  \Doi{10.1088/0953-8984/20/21/215201}{\bibinfo {journal} {Journal of Physics
  Condensed Matter}}\ }%
  \textbf{\bibinfo {volume} {20}},\ \bibinfo {pages} {215201} (\bibinfo {year}
  {2008})%
  \bibAnnoteFile{NoStop}{nilton:feru}%
\bibitem{artigo:paduani}%
  \BibitemOpen
  \bibfield{author}{%
  \bibinfo {author} {\bibfnamefont{W.}~\bibnamefont{P\"{o}ttker}}, \bibinfo
  {author} {\bibfnamefont{C.}~\bibnamefont{Paduani}}, \bibinfo {author}
  {\bibfnamefont{J.}~\bibnamefont{Ardisson}},\ and\ \bibinfo {author}
  {\bibfnamefont{M.}~\bibnamefont{Ioshida}},\ }%
  \bibfield{journal}{%
  \bibinfo {journal} {Phys. Stat. Sol. B}\ }%
  \textbf{\bibinfo {volume} {241}},\ \bibinfo {pages} {2586} (\bibinfo {year}
  {2004})%
  \bibAnnoteFile{NoStop}{artigo:paduani}%
\bibitem{kikuchi1980}%
  \BibitemOpen
  \bibfield{author}{%
  \bibinfo {author} {\bibfnamefont{R.}~\bibnamefont{Kikuchi}}, \bibinfo
  {author} {\bibfnamefont{J.}~\bibnamefont{Sanchez}}, \bibinfo {author}
  {\bibfnamefont{D.}~\bibnamefont{De~Fontaine}},\ and\ \bibinfo {author}
  {\bibfnamefont{H.}~\bibnamefont{Yamauchi}},\ }%
  \bibfield{journal}{%
  \bibinfo {journal} {Acta Metallurgica}\ }%
  \textbf{\bibinfo {volume} {28}},\ \bibinfo {pages} {651} (\bibinfo {year}
  {1980})%
  \bibAnnoteFile{NoStop}{kikuchi1980}%
\bibitem{livro:cardy}%
  \BibitemOpen
  \bibfield{author}{%
  \bibinfo {author} {\bibfnamefont{J.}~\bibnamefont{Cardy}},\ }%
  \emph{\bibinfo {title} {{Scaling and Renormalization in Statistical
  Physics}}}\ (\bibinfo {publisher} {Cambridge University Press},\ \bibinfo
  {address} {Cambridge, UK},\ \bibinfo {year} {1996})%
  \bibAnnoteFile{NoStop}{livro:cardy}%
\bibitem{belanger2000}%
  \BibitemOpen
  \bibfield{author}{%
  \bibinfo {author} {\bibfnamefont{D.}~\bibnamefont{Belanger}},\ }%
  \bibfield{journal}{%
  \bibinfo {journal} {Braz. J. Phys.}\ }%
  \textbf{\bibinfo {volume} {30}},\ \bibinfo {pages} {682} (\bibinfo {year}
  {2000})%
  \bibAnnoteFile{NoStop}{belanger2000}%
\bibitem{hasenbusch2007uc}%
  \BibitemOpen
  \bibfield{author}{%
  \bibinfo {author} {\bibfnamefont{M.}~\bibnamefont{Hasenbusch}}, \bibinfo
  {author} {\bibfnamefont{F.}~\bibnamefont{Toldin}}, \bibinfo {author}
  {\bibfnamefont{A.}~\bibnamefont{Pelissetto}},\ and\ \bibinfo {author}
  {\bibfnamefont{E.}~\bibnamefont{Vicari}},\ }%
  \bibfield{journal}{%
  \bibinfo {journal} {J. Stat. Mech.}\ }%
  \textbf{\bibinfo {volume} {2007}},\ \bibinfo {pages} {P02016} (\bibinfo
  {year} {2007})%
  \bibAnnoteFile{NoStop}{hasenbusch2007uc}%
\bibitem{livro:barkema}%
  \BibitemOpen
  \bibfield{author}{%
  \bibinfo {author} {\bibfnamefont{M.~E.~J.}\ \bibnamefont{Newman}}\ and\
  \bibinfo {author} {\bibfnamefont{G.~T.}\ \bibnamefont{Barkema}},\ }%
  \emph{\bibinfo {title} {\uppercase{M}onte \uppercase{C}arlo Methods in
  Statistical Physics}}\ (\bibinfo {publisher} {Oxford University Press},\
  \bibinfo {address} {New York, USA},\ \bibinfo {year} {1999})%
  \bibAnnoteFile{NoStop}{livro:barkema}%
\bibitem{ballesteros1998}%
  \BibitemOpen
  \bibfield{author}{%
  \bibinfo {author} {\bibfnamefont{H.~G.}\ \bibnamefont{Ballesteros}}, \bibinfo
  {author} {\bibfnamefont{L.~A.}\ \bibnamefont{Fern\'andez}}, \bibinfo {author}
  {\bibfnamefont{V.}~\bibnamefont{Mart\'in-Mayor}}, \bibinfo {author}
  {\bibfnamefont{A.}~\bibnamefont{Mu\~noz Sudupe}}, \bibinfo {author}
  {\bibfnamefont{G.}~\bibnamefont{Parisi}},\ and\ \bibinfo {author}
  {\bibfnamefont{J.~J.}\ \bibnamefont{Ruiz-Lorenzo}},\ }%
  \bibfield{journal}{%
  \Doi{10.1103/PhysRevB.58.2740}{\bibinfo {journal} {Phys. Rev. B}}\ }%
  \textbf{\bibinfo {volume} {58}},\ \bibinfo {pages} {2740} (\bibinfo {year}
  {1998})%
  \bibAnnoteFile{NoStop}{ballesteros1998}%
\bibitem{calabrese2003tdr}%
  \BibitemOpen
  \bibfield{author}{%
  \bibinfo {author} {\bibfnamefont{P.}~\bibnamefont{Calabrese}}, \bibinfo
  {author} {\bibfnamefont{V.}~\bibnamefont{Mart{\'\i}n-Mayor}}, \bibinfo
  {author} {\bibfnamefont{A.}~\bibnamefont{Pelissetto}},\ and\ \bibinfo
  {author} {\bibfnamefont{E.}~\bibnamefont{Vicari}},\ }%
  \bibfield{journal}{%
  \bibinfo {journal} {Phys. Rev. E}\ }%
  \textbf{\bibinfo {volume} {68}},\ \bibinfo {pages} {36136} (\bibinfo {year}
  {2003})%
  \bibAnnoteFile{NoStop}{calabrese2003tdr}%
\bibitem{hasenbusch2007+-J}%
  \BibitemOpen
  \bibfield{author}{%
  \bibinfo {author} {\bibfnamefont{M.}~\bibnamefont{Hasenbusch}}, \bibinfo
  {author} {\bibfnamefont{F.}~\bibnamefont{Toldin}}, \bibinfo {author}
  {\bibfnamefont{A.}~\bibnamefont{Pelissetto}},\ and\ \bibinfo {author}
  {\bibfnamefont{E.}~\bibnamefont{Vicari}},\ }%
  \bibfield{journal}{%
  \bibinfo {journal} {Phys. Rev. B}\ }%
  \textbf{\bibinfo {volume} {76}},\ \bibinfo {pages} {94402} (\bibinfo {year}
  {2007})%
  \bibAnnoteFile{NoStop}{hasenbusch2007+-J}%
\bibitem{livro:callen}%
  \BibitemOpen
  \bibfield{author}{%
  \bibinfo {author} {\bibfnamefont{H.~B.}\ \bibnamefont{Callen}},\ }%
  \emph{\bibinfo {title} {Thermodynamics and an Introduction to
  Thermostatistics}}\ (\bibinfo {publisher} {John Wiley \& Sons},\ \bibinfo
  {address} {New York, USA},\ \bibinfo {year} {1985})%
  \bibAnnoteFile{NoStop}{livro:callen}%
\bibitem{livro:julia}%
  \BibitemOpen
  \bibfield{author}{%
  \bibinfo {author} {\bibfnamefont{J.}~\bibnamefont{Yeomans}},\ }%
  \emph{\bibinfo {title} {Statistcal Mechanics of Phase Transtions}}\ (\bibinfo
  {publisher} {Clarendon Press},\ \bibinfo {address} {New York, USA},\ \bibinfo
  {year} {1992})%
  \bibAnnoteFile{NoStop}{livro:julia}%
\bibitem{artigo:metropolis}%
  \BibitemOpen
  \bibfield{author}{%
  \bibinfo {author} {\bibfnamefont{N.}~\bibnamefont{Metropolis}}, \bibinfo
  {author} {\bibfnamefont{A.}~\bibnamefont{Rosenbluth}}, \bibinfo {author}
  {\bibfnamefont{M.}~\bibnamefont{Rosenbluth}}, \bibinfo {author}
  {\bibfnamefont{A.}~\bibnamefont{Teller}},\ and\ \bibinfo {author}
  {\bibfnamefont{E.}~\bibnamefont{Teller}},\ }%
  \bibfield{journal}{%
  \bibinfo {journal} {J. Chem. Phys.}\ }%
  \textbf{\bibinfo {volume} {21}},\ \bibinfo {pages} {1087} (\bibinfo {year}
  {1953})%
  \bibAnnoteFile{NoStop}{artigo:metropolis}%
\bibitem{artigo:shift-register}%
  \BibitemOpen
  \bibfield{author}{%
  \bibinfo {author} {\bibfnamefont{S.}~\bibnamefont{Kirkpatrick}},\ }%
  \bibfield{journal}{%
  \Doi{10.1016/0021-9991(81)90227-8}{\bibinfo {journal} {J. Comput. Phys.}}\ }%
  \textbf{\bibinfo {volume} {40}},\ \bibinfo {pages} {517} (\bibinfo {year}
  {1981})%
  \bibAnnoteFile{NoStop}{artigo:shift-register}%
\bibitem{artigo:landau}%
  \BibitemOpen
  \bibfield{author}{%
  \bibinfo {author} {\bibfnamefont{A.~M.}\ \bibnamefont{Ferrenberg}}\ and\
  \bibinfo {author} {\bibfnamefont{D.~P.}\ \bibnamefont{Landau}},\ }%
  \bibfield{journal}{%
  \Doi{10.1103/PhysRevB.44.5081}{\bibinfo {journal} {Phys. Rev. B}}\ }%
  \textbf{\bibinfo {volume} {44}},\ \bibinfo {pages} {5081} (\bibinfo {year}
  {1991})%
  \bibAnnoteFile{NoStop}{artigo:landau}%
\bibitem{shan-ho:private}%
  \BibitemOpen
  \bibfield{author}{%
  \bibinfo {author} {\bibfnamefont{S.-H.}\ \bibnamefont{Tsai}},\ }%
  \bibinfo {howpublished} {private communication} (\bibinfo {year} {2008})%
  \bibAnnoteFile{NoStop}{shan-ho:private}%
\bibitem{livro:calculo-numerico}%
  \BibitemOpen
  \bibfield{author}{%
  \bibinfo {author} {\bibfnamefont{K.~E.}\ \bibnamefont{Atkinson}},\ }%
  \emph{\bibinfo {title} {An Introduction to Numerical Analysis}},\ \bibinfo
  {edition} {2nd}\ ed.\ (\bibinfo {publisher} {John Wiley \& Sons},\ \bibinfo
  {address} {New York, USA},\ \bibinfo {year} {1989})\ ISBN \bibinfo {isbn}
  {9780471500230}%
  \bibAnnoteFile{NoStop}{livro:calculo-numerico}%
\bibitem{artigo:ferrenberg:histograma1}%
  \BibitemOpen
  \bibfield{author}{%
  \bibinfo {author} {\bibfnamefont{A.~M.}\ \bibnamefont{Ferrenberg}}\ and\
  \bibinfo {author} {\bibfnamefont{R.~H.}\ \bibnamefont{Swendsen}},\ }%
  \bibfield{journal}{%
  \Doi{10.1103/PhysRevLett.61.2635}{\bibinfo {journal} {Phys. Rev. Lett.}}\ }%
  \textbf{\bibinfo {volume} {61}},\ \bibinfo {pages} {2635} (\bibinfo {year}
  {1988})%
  \bibAnnoteFile{NoStop}{artigo:ferrenberg:histograma1}%
\bibitem{artigo:ferrenberg:histograma2}%
  \BibitemOpen
  \bibfield{author}{%
  \bibinfo {author} {\bibfnamefont{A.~M.}\ \bibnamefont{Ferrenberg}}\ and\
  \bibinfo {author} {\bibfnamefont{R.~H.}\ \bibnamefont{Swendsen}},\ }%
  \bibfield{journal}{%
  \Doi{10.1103/PhysRevLett.63.1195}{\bibinfo {journal} {Phys. Rev. Lett.}}\ }%
  \textbf{\bibinfo {volume} {63}},\ \bibinfo {pages} {1195} (\bibinfo {year}
  {1989})%
  \bibAnnoteFile{NoStop}{artigo:ferrenberg:histograma2}%
\bibitem{artigo:butera}%
  \BibitemOpen
  \bibfield{author}{%
  \bibinfo {author} {\bibfnamefont{P.}~\bibnamefont{Butera}}\ and\ \bibinfo
  {author} {\bibfnamefont{M.}~\bibnamefont{Comi}},\ }%
  \bibfield{journal}{%
  \Doi{10.1103/PhysRevB.62.14837}{\bibinfo {journal} {Phys. Rev. B}}\ }%
  \textbf{\bibinfo {volume} {62}},\ \bibinfo {pages} {14837} (\bibinfo {year}
  {2000})%
  \bibAnnoteFile{NoStop}{artigo:butera}%
\bibitem{livro:murilinho}%
  \BibitemOpen
  \bibfield{author}{%
  \bibinfo {author} {\bibfnamefont{P.~M.~C.}\ \bibnamefont{de~Oliveira}},\ }%
  \emph{\bibinfo {title} {{Computing Boolean Statistical Models}}}\ (\bibinfo
  {publisher} {World Scientific},\ \bibinfo {address} {Singapore},\ \bibinfo
  {year} {1991})%
  \bibAnnoteFile{NoStop}{livro:murilinho}%
\bibitem{artigo:landau-z}%
  \BibitemOpen
  \bibfield{author}{%
  \bibinfo {author} {\bibfnamefont{S.}~\bibnamefont{Wansleben}}\ and\ \bibinfo
  {author} {\bibfnamefont{D.}~\bibnamefont{Landau}},\ }%
  \bibfield{journal}{%
  \bibinfo {journal} {Physical Review B}\ }%
  \textbf{\bibinfo {volume} {43}},\ \bibinfo {pages} {6006} (\bibinfo {year}
  {1991})%
  \bibAnnoteFile{NoStop}{artigo:landau-z}%
\bibitem{compostrini1999}%
  \BibitemOpen
  \bibfield{author}{%
  \bibinfo {author} {\bibfnamefont{M.}~\bibnamefont{Campostrini}}, \bibinfo
  {author} {\bibfnamefont{A.}~\bibnamefont{Pelissetto}}, \bibinfo {author}
  {\bibfnamefont{P.}~\bibnamefont{Rossi}},\ and\ \bibinfo {author}
  {\bibfnamefont{E.}~\bibnamefont{Vicari}},\ }%
  \bibfield{journal}{%
  \Doi{10.1103/PhysRevE.60.3526}{\bibinfo {journal} {Phys. Rev. E}}\ }%
  \textbf{\bibinfo {volume} {60}},\ \bibinfo {pages} {3526} (\bibinfo {year}
  {1999})%
  \bibAnnoteFile{NoStop}{compostrini1999}%
\bibitem{guillou-zinn1980}%
  \BibitemOpen
  \bibfield{author}{%
  \bibinfo {author} {\bibfnamefont{J.~C.}\ \bibnamefont{{Le Guillou}}}\ and\
  \bibinfo {author} {\bibfnamefont{J.}~\bibnamefont{Zinn-Justin}},\ }%
  \bibfield{journal}{%
  \Doi{10.1103/PhysRevB.21.3976}{\bibinfo {journal} {Phys. Rev. B}}\ }%
  \textbf{\bibinfo {volume} {21}},\ \bibinfo {pages} {3976} (\bibinfo {year}
  {1980})%
  \bibAnnoteFile{NoStop}{guillou-zinn1980}%
\bibitem{guillou-zinn1987}%
  \BibitemOpen
  \bibfield{author}{%
  \bibinfo {author} {\bibfnamefont{J.~C.}\ \bibnamefont{{Le Guillou}}}\ and\
  \bibinfo {author} {\bibfnamefont{J.}~\bibnamefont{Zinn-Justin}},\ }%
  \bibfield{journal}{%
  \bibinfo {journal} {Journal de Physique}\ }%
  \textbf{\bibinfo {volume} {48}},\ \bibinfo {pages} {19} (\bibinfo {year}
  {1987})%
  \bibAnnoteFile{NoStop}{guillou-zinn1987}%
\bibitem{lundow2009}%
  \BibitemOpen
  \bibfield{author}{%
  \bibinfo {author} {\bibfnamefont{P.}~\bibnamefont{Lundow}}, \bibinfo {author}
  {\bibfnamefont{K.}~\bibnamefont{Markström}},\ and\ \bibinfo {author}
  {\bibfnamefont{A.}~\bibnamefont{Rosengren}},\ }%
  \bibfield{journal}{%
  \bibinfo {journal} {Philosophical Magazine}\ }%
  \textbf{\bibinfo {volume} {89}},\ \bibinfo {pages} {2009} (\bibinfo {year}
  {2009})%
  \bibAnnoteFile{NoStop}{lundow2009}%
\bibitem{kaul1983}%
  \BibitemOpen
  \bibfield{author}{%
  \bibinfo {author} {\bibfnamefont{S.}~\bibnamefont{Kaul}},\ }%
  \bibfield{journal}{%
  \bibinfo {journal} {Phys. Rev. B}\ }%
  \textbf{\bibinfo {volume} {27}},\ \bibinfo {pages} {6923} (\bibinfo {year}
  {1983})%
  \bibAnnoteFile{NoStop}{kaul1983}%
\end{thebibliography}
\end{document}